\documentclass[10pt]{revtex4-2}
\usepackage[T1]{fontenc}
\usepackage[utf8]{inputenc}
\usepackage{fancybox}
\usepackage{calc}
\usepackage{amsmath}
\usepackage{amssymb}
\usepackage{amsthm}
\usepackage{stmaryrd}
\usepackage{graphicx}
\PassOptionsToPackage{version=3}{mhchem}
\usepackage{mhchem}
\usepackage{color}
\usepackage{xcolor}

\usepackage{boxedminipage}
\makeatletter
\usepackage{enumitem}

\usepackage{physics}
\usepackage{bm}
\usepackage{tcolorbox}
\usepackage{hyperref}
\usepackage{booktabs}

\usepackage[normalem]{ulem} 
\usepackage{xcolor}

\newcommand{\e}{\mathrm{e}}

\newcommand{\bg}{\bar{g}}
\newcommand{\bR}{\bar{R}}

\newcommand{\bGamma}{\bar{\Gamma}}

\@ifundefined{date}{}{\date{}}

\usepackage{amsfonts}
\usepackage[export]{adjustbox}
\graphicspath{ {./images/} }

\makeatother

\begin{document}

\title{\normalsize Static Tidal Perturbations of Relativistic Stars: Corrected Center Expansion and Love Numbers-I}
\author{{\normalsize{}{}{}{}{}{}{}{}{}{}{}{}{}{}{}Emel Altas}}
\email{emel.altas@agu.edu.tr}

\affiliation{Department of Engineering Sciences, Abdullah Gul University, 38080 Kayseri, Turkey}
\author{{\normalsize{}{}{}{}{}{}{}{}{}{}{}{}{}{}{}Ercan Kilicarslan}}
\email{ercan.kilicarslan@usak.edu.tr}

\affiliation{Department of Mathematics, Usak University, 64200, Usak, Turkey}

\author{{\normalsize{}{}{}{}{}{}{}{}{}{}{}{}{}{}{}Onur Oktay}}
\email{oktay24005@gmail.com}

\affiliation{Department of Engineering Sciences, Abdullah Gul University, 38080 Kayseri, Turkey}

\author{{\normalsize{}{}{}{}{}{}{}{}{}{}{}{}{}{}{}Bayram Tekin}}
\email{bayram.tekin@bilkent.edu.tr}

\affiliation{Department of Physics, Bilkent University, 06800, Ankara,
Turkey}
\date{{\normalsize{}{}{}{}{}{}{}{}{{}{}{}{}\today}}}
\begin{abstract}
We revisit static tidal perturbations of relativistic stars with emphasis on two technical issues in the standard quadrupolar formulation. First, we derive the regular-center Frobenius expansion of the interior even-parity master function and obtain a corrected subleading coefficient, which differs from the expression commonly used in the literature. Second, we derive the static even-parity master equation on a Schwarzschild--de Sitter background, extending the usual asymptotically flat problem to a two-horizon geometry. To place these results on a common footing, we also show how the general interior even-parity system in Regge--Wheeler gauge reduces to the standard quadrupolar equation used in Love-number calculations. Numerical integrations for polytropic equations of state show that the corrected center coefficient affects only subleading initial data and leaves the extracted Love number $k_2$ unchanged within numerical accuracy. Taken together, these results fix the regular-center input to the standard quadrupolar problem and extend the static even-parity formalism to Schwarzschild–de Sitter backgrounds.
\end{abstract}
\maketitle

\section{Introduction}

In Newtonian gravity, the deformation of a self-gravitating body placed in an external field is characterized by a set of dimensionless response coefficients known as the Love numbers. These quantities measure how an applied tidal field induces multipole moments in the body and thereby encode, in a compact way, the body's internal rigidity and density distribution. In this sense, Love numbers provide a direct link between internal structure and observable long-range fields. Their relativistic counterparts have become especially important in compact-star physics, where they furnish one of the cleanest probes of finite-size effects in strongly gravitating systems \cite{Hinderer2008,FlanaganHinderer2008,BinningtonPoisson2009,DamourNagar2009}; for recent broad reviews, see Refs.~\cite{Chakraborty:2026qru,Rodriguez:2026iot}.
In this paper, we focus on three technical points within this framework: a corrected regular-center expansion for the quadrupolar interior equation, a controlled numerical test of its impact on $k_2$ and the corresponding static even-parity equation on Schwarzschild--de Sitter.

The Newtonian gravitational potential $U$ satisfies Poisson's equation,
\begin{align}
	\nabla^{2}U = 4\pi G\,\rho \,,
\end{align}
where $\rho$ is the mass density. In vacuum, $\nabla^{2}U=0$, and for an isolated body of mass $M$ one recovers
\begin{align}
	U=-\frac{GM}{r}\, .
\end{align}
When the body is immersed in the weak tidal field of a distant companion, spherical symmetry is broken, and the potential admits a multipolar expansion. In the static or adiabatic regime, in which the external field changes on a timescale much longer than the body's internal dynamical time, one may write
\begin{align}
	U = -\frac{GM}{r} - \sum_{l\geq 2}\frac{1}{l(l-1)}
	\left(1+2k_{l}\left(\frac{R}{r}\right)^{2l+1}\right)
	\mathcal{E}_{L}x^{L},
\end{align}
where $R$ is the body's radius, $k_{l}$ are the Newtonian Love numbers, $L$ denotes a multi-index, $\mathcal{E}_{L}$ is the symmetric trace-free tidal moment of multipole order $l$, and $x^{L}$ denotes the corresponding Cartesian multipole. The term proportional to $\mathcal{E}_{L}x^{L}$ describes the applied external tidal field, while the term proportional to $k_l$ describes the induced response of the body. Thus, the Love numbers quantify the degree to which a self-gravitating object departs from rigid behavior under tidal forcing.

In general relativity, the Newtonian potential is replaced by the spacetime metric, and tidal deformability is determined by solving the Einstein equations for a perturbed gravitating configuration. The field equations are
\begin{align}
\label{EinsteinEq}
	\bar G_{\mu\nu} = 8\pi \bar T_{\mu\nu} \,,
\end{align}
where $\bar T_{\mu\nu}$ denotes the stress-energy tensor and $\bar G_{\mu\nu}$ is the Einstein tensor,
\begin{align}
\label{EinsteinTensor}
	\bar G_{\mu\nu} = \bar R_{\mu\nu} -\frac{1}{2}\bar g_{\mu\nu}\bar R \,,
\end{align}
and we set $G=c=1$ throughout. For a static, spherically symmetric compact star, the unperturbed spacetime is described by a perfect-fluid solution of the Tolman–Oppenheimer–Volkoff equations. Once an external tidal field is introduced, one perturbs the metric according to
\begin{align}
	g_{\mu\nu}=\bar g_{\mu\nu}+h_{\mu\nu},
\end{align}
which solves the linearized Einstein equations for the perturbation $h_{\mu\nu}$. The tidal response is then extracted from the asymptotic behavior of the exterior solution \cite{Hinderer2008,BinningtonPoisson2009,DamourNagar2009}.

The relativistic problem is richer than its Newtonian counterpart in several respects. First, the perturbation naturally decomposes into even and odd-parity sectors~\cite{Zerilli1970}, often referred to as the electric and magnetic-type tidal sectors, respectively. The even-parity sector is the direct relativistic analogue of the Newtonian tidal deformation and is associated with induced mass multipole moments. The odd-parity sector has no Newtonian counterpart and describes genuinely relativistic aspects of the response, related to current-type or gravitomagnetic deformations \cite{BinningtonPoisson2009,DamourNagar2009}. Second, the relativistic field equations must be handled with care with respect to gauge choice, regularity, and matching conditions between the stellar interior and the exterior vacuum region. Third, the magnetic-type response contains subtleties related to the fluid state in the zero-frequency limit, a point that has been clarified in the subsequent literature \cite{PaniEtAl2018}.

The modern relativistic theory of Love numbers was developed in a sequence of influential works. Flanagan and Hinderer showed that the tidal deformability of neutron stars leaves a measurable imprint on the phase of the inspiral waveform and can therefore be constrained by gravitational-wave observations \cite{FlanaganHinderer2008}. Hinderer then gave a practical formulation of the static quadrupolar problem for relativistic stars, reducing the even-parity sector to a single second-order differential equation whose solution determines the quadrupolar Love number $k_2$ \cite{Hinderer2008}. Soon afterward, Binnington and Poisson formulated a general relativistic theory of electric- and magnetic-type Love numbers for compact objects and emphasized the precise identification of the growing and decaying pieces of the exterior solution \cite{BinningtonPoisson2009}. In parallel, Damour and Nagar analyzed several relativistic tidal coefficients and highlighted their strong dependence on stellar compactness \cite{DamourNagar2009}. Together, these works established the basic relativistic framework now used in studies of neutron-star tidal deformability.

The astrophysical importance of this framework has increased dramatically in the era of gravitational-wave astronomy. The binary neutron-star event GW170817 provided the first direct observation of a neutron-star inspiral in gravitational waves \cite{GW170817}. Subsequent analyzes of that event improved the bounds on the binary's tidal parameters and showed that tidal deformability can be used to rule out classes of equations of state \cite{Abbott2018PropertiesGW170817}. For this reason, the relativistic Love numbers of compact stars are now recognized as central observables connecting strong-field gravity, nuclear matter, and gravitational-wave data analysis \cite{FlanaganHinderer2008,Hinderer2008,Abbott2018PropertiesGW170817}.

Despite this mature physical picture, several technical issues in the standard static treatment remain worth revisiting. The regular-center expansion used to initialize the interior even-parity equation is often quoted without a derivation, and the subleading coefficient commonly used in the literature is incorrect.
It is also not obvious a priori whether correcting that coefficient changes the extracted Love number in a measurable way.
The situation is particularly delicate in the even-parity sector, where one ultimately reduces the full perturbation problem to the familiar master equation for the metric function $H(r)$ introduced by Hinderer \cite{Hinderer2008}.
Here, that reduction is included because it is the technical bridge to the corrected center expansion and to the numerical comparison.
In addition, the static even-parity problem is usually formulated only for asymptotically flat exteriors, so the corresponding Schwarzschild–de Sitter equation is not part of the standard formulation. These questions motivate the present analysis.

One purpose of the present work is therefore to present the construction of relativistic Love numbers in a fully explicit form, starting from the linearized field equations and carrying out the decomposition into scalar, vector, and tensor spherical harmonics in detail. We work in Regge--Wheeler gauge, which remains the natural framework for separating the even- and odd-parity sectors of static perturbations \cite{ReggeWheeler1957}. In the exterior vacuum region, we derive the radial equations for both parity sectors and identify the growing and decaying branches from which the electric- and magnetic-type Love numbers are extracted. In the stellar interior, we explicitly include the fluid perturbations and show how the perturbation system reduces to the master equations relevant to tidal response.

A second goal of this paper is to revisit the regular-center analysis of the even-parity interior equation. The behavior of the master field near the stellar center is usually imposed through a Frobenius expansion, but the detailed derivation of the subleading coefficients is not always displayed. Here, we perform that expansion carefully and obtain a corrected form of the near-center coefficient in the regular solution for the quadrupolar master function. This correction is analytically important because it clarifies the exact regular series structure of the interior solution. At the same time, our numerical analysis shows that, for the polytropic models considered here, the correction does not lead to a visible change in the final value of the Love number $k_2$, since it contributes only through subleading powers in the initial data.

A third aim is to extend the static even-parity analysis beyond asymptotically flat spacetimes. When a positive cosmological constant is present, the natural spherically symmetric vacuum background is the Schwarzschild-de Sitter geometry, which possesses both a black-hole horizon and a cosmological horizon. In such a setting, the usual asymptotic analysis at spatial infinity is replaced by a two-horizon problem, and the notion of tidal response must be reexamined accordingly. For this reason, we derive the corresponding static even-parity master equation on the Schwarzschild–de Sitter background. This furnishes a natural de Sitter extension of the standard static tidal problem. 

The quadrupolar electric-type Love number is of particular interest since it is the quantity most directly used in gravitational-wave applications. It is customary to package the quadrupolar response in terms of the tidal deformability parameter $\lambda_t$ and the dimensionless deformability
\begin{align}
	\bar\Lambda = \frac{\lambda_t}{M^5} = \frac{2}{3} k_2 C^{-5},
\end{align}
where
\begin{align}
	C=\frac{M}{R},
\end{align}
is the stellar compactness. Due to the strong $C^{-5}$ factor, even modest changes in compactness 
-- equivalently, at fixed mass, modest changes in the radius -- can produce large changes in $\bar\Lambda$.
This sensitivity is precisely what makes tidal deformability such a useful observable for constraining the equation of state of neutron-star matter \cite{FlanaganHinderer2008,Hinderer2008,Abbott2018PropertiesGW170817}.

The paper is organized as follows. In Sec.~\ref{sec:Unperturbed_Configuration} we review the unperturbed, static, spherically symmetric perfect-fluid configuration and collect the relevant background relations. In Sec.~\ref{sec:perturbed_configuration} we formulate the linear perturbation problem, introduce the scalar, vector, and tensor spherical harmonics, and derive the odd- and even-parity perturbation equations in the exterior and interior regions. In Sec.~\ref{sec:hinderer_reduction} we show explicitly how the interior even-parity system reduces to the standard Hinderer equation for the quadrupolar response. In Sec.~\ref{sec:center_expansion} we perform the regular-center Frobenius analysis and derive the corrected near-center expansion of the master function. In Sec.~\ref{sec:exterior_solution} we discuss the exterior solution and the extraction of the Love number by matching. In Sec.~\ref{sec:numerical_results} we present numerical results for polytropic stellar models and compare the original and corrected near-center prescriptions. In Sec.~\ref{sec:sds_master} we extend the static even-parity analysis to the Schwarzschild--de Sitter background and derive the corresponding master equation. We conclude in Sec.~\ref{sec:conclusions}.

Throughout this work, our emphasis is as much on clarity as on results. We aim to provide a transparent and self-contained derivation of relativistic tidal perturbation theory in a form that makes the intermediate steps explicit, while also clarifying several technical points that are frequently passed over quickly in the literature: the full reduction to the Hinderer equation, the detailed structure of the regular-center expansion, and the extension of the static tidal problem to a spacetime with a cosmological horizon.

\section{Unperturbed Configuration}
\label{sec:Unperturbed_Configuration}

In this section, we summarize the static, spherically symmetric perfect-fluid background that models an isolated relativistic star. We introduce the background metric and matter variables and collect the basic relations that will serve as the starting point for the tidal perturbation analysis.

In the absence of external sources, a static perfect-fluid body settles into a spherically symmetric equilibrium configuration. We therefore consider an isolated, static, spherically symmetric distribution of perfect fluid. In Schwarzschild-like coordinates, the background line element may be written as
\begin{align}
\label{bg_metric}
	ds^{2}
	=
	\bar g_{\mu\nu}dx^{\mu}dx^{\nu}
	=
	-e^{2\psi(r)}dt^{2}
	+
	f^{-1}(r)\,dr^{2}
	+
	r^{2}\left(d\theta^{2}+\sin^{2}\theta\,d\phi^{2}\right),
\end{align}
where the functions $\psi(r)$ and $f(r)$ are determined by the background Einstein equations together with the fluid equation of state.
For later use, we recall that the Levi--Civita connection associated with the background metric is
\begin{align}
\label{bg_connection}
	\bar\Gamma^{\mu}{}_{\nu\sigma}
	=
	\frac{1}{2}\bar g^{\mu\lambda}
	\left(
	\partial_{\nu}\bar g_{\sigma\lambda}
	+
	\partial_{\sigma}\bar g_{\nu\lambda}
	-
	\partial_{\lambda}\bar g_{\nu\sigma}
	\right),
\end{align}
and the corresponding Ricci tensor is
\begin{align}
\label{bg_Ricci}
	\bar R_{\mu\nu}
	=
	\partial_{\sigma}\bar\Gamma^{\sigma}{}_{\mu\nu}
	-
	\partial_{\nu}\bar\Gamma^{\sigma}{}_{\sigma\mu}
	+
	\bar\Gamma^{\sigma}{}_{\sigma\lambda}\bar\Gamma^{\lambda}{}_{\mu\nu}
	-
	\bar\Gamma^{\sigma}{}_{\nu\lambda}\bar\Gamma^{\lambda}{}_{\mu\sigma}.
\end{align}
For completeness, the nonvanishing connection coefficients and curvature components associated with Eq.~\eqref{bg_metric} are collected in Appendix~\ref{app:background_geometry}.

The matter source is considered to be a perfect fluid with a stress-energy tensor.
\begin{align}
\label{perfect_fluid_Tmunu}
	\bar T_{\mu\nu}
	=
	(\bar\rho+\bar p)\bar u_{\mu}\bar u_{\nu}
	+
	\bar p\,\bar g_{\mu\nu},
\end{align}
where $\bar\rho(r)$ and $\bar p(r)$ denote the background energy density and pressure, respectively, and $\bar u^{\mu}$ is the fluid's four-velocity. Since the background is static, the fluid is at rest in these coordinates, so its four-velocity is purely timelike. The normalization condition
\begin{align}
	\bar u_{\mu}\bar u^{\mu}=-1,
\end{align}
then gives
\begin{align}
\label{four_vels}
	\bar u^{\mu}
	=
	\left(e^{-\psi},0,0,0\right),
	\qquad
	\bar u_{\mu}
	=
	\left(-e^{\psi},0,0,0\right).
\end{align}
Substituting Eq.~\eqref{four_vels} into Eq.~\eqref{perfect_fluid_Tmunu}, one finds the nonvanishing background stress-energy components:
\begin{align}
\label{bg_T_components}
	\bar T_{tt}
	=
	e^{2\psi}\bar\rho,
	\qquad
	\bar T_{rr}
	=
	f^{-1}\bar p,
	\qquad
	\bar T_{\theta\theta}
	=
	r^{2}\bar p,
	\qquad
	\bar T_{\phi\phi}
	=
	r^{2}\sin^{2}\theta\,\bar p.
\end{align}
The background configuration is governed by the Einstein equations together with stress-energy conservation. In particular, the covariant conservation law
\begin{align}
\label{bg_conservation}
	\bar\nabla_{\mu}\bar T^{\mu\nu}=0,
\end{align}
yields the relativistic condition for hydrostatic equilibrium. Evaluating the radial component of Eq.~\eqref{bg_conservation}, one obtains
\begin{align}
\label{TOV}
	\partial_{r}\bar p
	=
	-(\bar\rho+\bar p)\,\partial_{r}\psi,
\end{align}
which is the Tolman–Oppenheimer–Volkoff equation in the form appropriate to the metric ansatz \eqref{bg_metric}. This relation expresses the balance between the inward gravitational pull encoded in the redshift function $\psi(r)$ and the outward pressure gradient of the fluid.

In the following subsections, we determine the explicit forms of the background solution in the exterior and interior regions and record the matching conditions at the stellar surface that will be needed for the perturbative analysis.

\subsection{Exterior Region Solution of the Unperturbed Configuration}
\label{subsec:exterior_unperturbed}

For $r>R$, the spacetime is a vacuum, and hence the stress-energy tensor vanishes:
\begin{align}
	\bar T_{\mu\nu}=0.
\end{align}
The background Einstein equations, therefore, reduce to
\begin{align}
\label{vacuum_EFE}
	\bar R_{\mu\nu}
	-\frac{1}{2}\bar g_{\mu\nu}\bar R
	=0.
\end{align}
Taking the trace of Eq.~\eqref{vacuum_EFE} immediately gives
\begin{align}
	\bar R=0,
\end{align}
and therefore
\begin{align}
	\bar R_{\mu\nu}=0.
\end{align}
Substituting the metric ansatz \eqref{bg_metric} into the vacuum field equations, and using the explicit Ricci components listed in Appendix~\ref{app:background_geometry}, one finds from the conditions $\bar R_{tt}=0$ and $\bar R_{rr}=0$ the first-order relation
\begin{align}
\label{psi_f_relation_vacuum}
	2\psi'(r)=\frac{f'(r)}{f(r)}.
\end{align}
Integrating Eq.~\eqref{psi_f_relation_vacuum}, one obtains
\begin{align}
	e^{2\psi(r)} = \mathcal{C}\, f(r),
\end{align}
where $\mathcal{C}$ is a constant of integration. Imposing asymptotic flatness, so that
\begin{align}
	e^{2\psi(r)}\to 1,
	\qquad
	f(r)\to 1
	\qquad
	(r\to\infty),
\end{align}
fixes $\mathcal{C}=1$. Thus,
\begin{align}
\label{psi_equals_f}
	f(r)=e^{2\psi(r)}.
\end{align}
The exterior metric therefore takes the Schwarzschild form,
\begin{align}
\label{Schwarzschild_sol}
	ds^{2}
	=
	-f(r)\,dt^{2}
	+
	f^{-1}(r)\,dr^{2}
	+
	r^{2}d\theta^{2}
	+
	r^{2}\sin^{2}\theta\,d\phi^{2},
\end{align}
with
\begin{align}
\label{Schwarzschild_f}
	f(r)=1-\frac{2M}{r},
\end{align}
where $M$ is an integration constant identified with the total gravitational mass of the star. Thus, outside the stellar surface, the background geometry is exactly Schwarzschild, as expected from Birkhoff's theorem.

\subsection{Interior Region Solution of the Unperturbed Configuration}
\label{subsec:interior_unperturbed}

For $r<R$, the matter source is described by the perfect-fluid stress-energy tensor \eqref{perfect_fluid_Tmunu}, and the background Einstein equations take the form
\begin{align}
\label{einstein_interior}
	\bar R_{\mu\nu}
	-\frac{1}{2}\bar g_{\mu\nu}\bar R
	=
	8\pi \bar T_{\mu\nu}.
\end{align}
Taking the trace of Eq.~\eqref{einstein_interior}, we obtain
\begin{align}
	\bar R=-8\pi \bar T,
\end{align}
where $\bar T=\bar g^{\mu\nu}\bar T_{\mu\nu}$ is the trace of the perfect-fluid stress-energy tensor. Using
\begin{align}
	\bar T=-\bar\rho+3\bar p,
\end{align}
it follows that
\begin{align}
\label{scalar_curvature_interior}
	\bar R=8\pi(\bar\rho-3\bar p).
\end{align}
It is convenient to introduce the enclosed mass function $m(r)$ through
\begin{align}
\label{mass_function_def}
	f(r)=1-\frac{2m(r)}{r}.
\end{align}
This function measures the total mass contained inside radius $r$, and matching to the exterior vacuum solution at the stellar surface requires
\begin{align}
	m(R)=M.
\end{align}
Substituting Eq.~\eqref{mass_function_def} into the $tt$ component of Eq.~\eqref{einstein_interior}, one finds the standard mass equation
\begin{align}
\label{mass_equation}
	\frac{dm}{dr}
	=
	4\pi r^{2}\bar\rho(r).
\end{align}
The remaining background equation may be obtained from the $rr$ component of the Einstein equations. Using the curvature expressions associated with the metric \eqref{bg_metric}, collected in Appendix~\ref{app:background_geometry}, one arrives at
\begin{align}
	\psi'(r)+\frac{1}{2r}-\frac{1}{2fr}
	=
	\frac{4\pi \bar p\,r}{f}.
\end{align}
Eliminating $f(r)$ in favor of the mass function using Eq.~\eqref{mass_function_def}, this relation can be written in the standard form
\begin{align}
\label{psi_prime}
	\psi'(r)
	=
	\frac{m(r)+4\pi r^{3}\bar p(r)}{r^{2}f(r)}.
\end{align}
Together with the hydrostatic equilibrium equation \eqref{TOV},
\begin{align}
	\bar p'(r)=-(\bar\rho+\bar p)\psi'(r),
\end{align}
Eqs.~\eqref{mass_equation} and \eqref{psi_prime} constitute the Tolman--Oppenheimer--Volkoff system governing the interior equilibrium of the relativistic star.
At the stellar surface $r=R$, the pressure vanishes,
\begin{align}
	\bar p(R)=0,
\end{align}
and the interior solution must be matched smoothly to the exterior Schwarzschild geometry. These background relations provide the unperturbed configuration about which the tidal perturbation problem will be developed.

\section{Perturbed Configuration}
\label{sec:perturbed_configuration}

We now consider a weak perturbation of the isolated, static background configuration, generated by a distant matter distribution, such as a companion located at a separation $a\gg R$. The characteristic timescale associated with variations of the external tidal field is of order $\sqrt{a^{3}/M}$, whereas the internal hydrodynamic response time of the star is of order $\sqrt{R^{3}/M}$. Since $a\gg R$, the external field varies much more slowly than the star responds internally. It is therefore consistent, in the adiabatic or static-tide regime, to treat the perturbation as time independent and to work only to first order in the perturbation amplitude.

Accordingly, we expand the spacetime metric about the background as
\begin{align}
	g_{\mu\nu}=\bar g_{\mu\nu}+h_{\mu\nu},
\end{align}
where $h_{\mu\nu}$ denotes the linear metric perturbation. Its trace is defined by
\begin{align}
	h \equiv \bar g^{\mu\nu}h_{\mu\nu}.
\end{align}
To first order in $h_{\mu\nu}$, the inverse metric is
\begin{align}
	g^{\mu\nu}=\bar g^{\mu\nu}-h^{\mu\nu}.
\end{align}
The perturbation is governed by the linearized Einstein equations,
\begin{align}
	\delta G_{\mu\nu}=8\pi\,\delta T_{\mu\nu},
\end{align}
where the linearized Einstein tensor is
\begin{align}
\label{linearized_einstein_tensor}
	\delta G_{\mu\nu}
	=
	\delta R_{\mu\nu}
	-\frac{1}{2}\bar g_{\mu\nu}\,\delta R
	-\frac{1}{2}h_{\mu\nu}\,\bar R.
\end{align}
Here $\delta R_{\mu\nu}$ and $\delta R$ denote, respectively, the linearized Ricci tensor and linearized scalar curvature. The linearized Ricci tensor may be written in covariant form as
\begin{align}
\label{deltaRmunu_covariant}
	\delta R_{\mu\nu}
	&=
	\bar\nabla_{\sigma}\delta\Gamma^{\sigma}{}_{\mu\nu}
	-
	\bar\nabla_{\nu}\delta\Gamma^{\sigma}{}_{\sigma\mu}
	\nonumber\\
	&=
	\frac{1}{2}
	\left(
	\bar\nabla_{\sigma}\bar\nabla_{\mu}h^{\sigma}{}_{\nu}
	+
	\bar\nabla_{\sigma}\bar\nabla_{\nu}h^{\sigma}{}_{\mu}
	-
	\bar\nabla_{\sigma}\bar\nabla^{\sigma}h_{\mu\nu}
	-
	\bar\nabla_{\mu}\bar\nabla_{\nu}h
	\right),
\end{align}
while the linearized scalar curvature is
\begin{align}
\label{deltaR_scalar}
	\delta R
	=
	\bar g^{\mu\nu}\delta R_{\mu\nu}
	-
	h^{\mu\nu}\bar R_{\mu\nu}
	=
	\bar\nabla_{\mu}\bar\nabla_{\nu}h^{\mu\nu}
	-
	\bar\nabla_{\mu}\bar\nabla^{\mu}h
	-
	h^{\mu\nu}\bar R_{\mu\nu}.
\end{align}

We now specialize to perturbations of the static, spherically symmetric background metric \eqref{bg_metric}, with $f(r)$ defined by Eq.~\eqref{mass_function_def}. In the static tidal limit, all background and perturbative quantities are taken to be time independent, so that
\begin{align}
	\partial_{t}h_{\mu\nu}=0,
	\qquad
	\partial_{t}\bar g_{\mu\nu}=0,
	\qquad
	\partial_{t}\bar\Gamma^{\sigma}{}_{\mu\nu}=0.
\end{align}
These conditions considerably simplify the structure of the linearized field equations and allow the perturbation problem to be reduced to a set of ordinary differential equations after separation of the angular dependence.
For later convenience, we introduce the angular Laplacian on the unit two-sphere,
\begin{align}
\label{D2_def}
	D^{2}
	\equiv
	\frac{1}{\sin\theta}\frac{\partial}{\partial\theta}
	\left(
	\sin\theta\,\frac{\partial}{\partial\theta}
	\right)
	+
	\frac{1}{\sin^{2}\theta}\frac{\partial^{2}}{\partial\phi^{2}}
	=
	\frac{\partial^{2}}{\partial\theta^{2}}
	+
	\cot\theta\,\frac{\partial}{\partial\theta}
	+
	\frac{1}{\sin^{2}\theta}\frac{\partial^{2}}{\partial\phi^{2}}.
\end{align}
This operator will appear repeatedly once the perturbations are decomposed into spherical harmonics.

Substituting the background connection coefficients, given in Appendix~\ref{app:background_geometry}, into the covariant expressions \eqref{deltaRmunu_covariant} and \eqref{deltaR_scalar}, and then imposing the static conditions above, yields explicit coordinate expressions for the components of $\delta R_{\mu\nu}$ in terms of partial derivatives of $h_{\mu\nu}$. Since these formulas are rather lengthy, we collect them in Appendices~\ref{app:linricci_diagonal} and \ref{app:linricci_offdiagonal}. They provide the starting point for the spherical-harmonic decomposition of the metric perturbation and for the subsequent reduction of the linearized field equations in both the exterior and interior regions.

\subsection{Scalar, Vector, and Tensor Spherical Harmonics}

To solve the linearized field equations, we decompose the metric perturbation into scalar, vector, and tensor spherical harmonics on the unit two-sphere. Our conventions are as follows~\cite{thorne1967non}:
\begin{itemize}
	\item Scalar spherical harmonics are denoted by $Y_{lm}(\theta,\phi)$.
	\item Even- and odd-parity vector spherical harmonics are denoted by $\psi_{A}^{lm}(\theta,\phi)$ and $\chi_{A}^{lm}(\theta,\phi)$, respectively.
	\item Even- and odd-parity tensor spherical harmonics are denoted by $\bar g_{AB}Y_{lm}(\theta,\phi)$, $\psi_{AB}^{lm}(\theta,\phi)$, and $\chi_{AB}^{lm}(\theta,\phi)$.
\end{itemize}
Here, lowercase Latin indices $a,b$ refer to the coordinates $(t,r)$, while uppercase Latin indices $A,B$ refer to the angular coordinates, $A,B\in\{\theta,\phi\}$.
The scalar spherical harmonics satisfy the eigenvalue equation
\begin{equation}
\label{scalar_sph_eigen}
	\left(D^{2}+l(l+1)\right)Y_{lm}(\theta,\phi)=0.
\end{equation}
The even-parity vector harmonics are defined by
\begin{align}
	\psi_{\theta}^{lm}(\theta,\phi)&=\partial_{\theta}Y_{lm}(\theta,\phi),
	&
	\psi_{\phi}^{lm}(\theta,\phi)&=\partial_{\phi}Y_{lm}(\theta,\phi),
\end{align}
and the odd-parity vector harmonics by
\begin{align}
	\chi_{\theta}^{lm}(\theta,\phi)&=-\frac{1}{\sin\theta}\partial_{\phi}Y_{lm}(\theta,\phi),
	&
	\chi_{\phi}^{lm}(\theta,\phi)&=\sin\theta\,\partial_{\theta}Y_{lm}(\theta,\phi).
\end{align}
The even-parity tensor harmonics are defined as
\begin{equation}
	\psi_{AB}^{lm}(\theta,\phi)
	=
	\left(
	\partial_{A}\partial_{B}
	+\frac{1}{2}l(l+1)\bar g_{AB}
	\right)Y_{lm}(\theta,\phi),
\end{equation}
while the odd-parity tensor harmonics are
\begin{equation}
	\chi_{AB}^{lm}(\theta,\phi)
	=
	-\frac{1}{2}
	\left(
	\varepsilon_{A}{}^{C}\partial_{B}
	+
	\varepsilon_{B}{}^{C}\partial_{A}
	\right)\partial_{C}Y_{lm}(\theta,\phi),
\end{equation}
where
\begin{equation}
	\varepsilon_{\theta}{}^{\phi}=\frac{1}{\sin\theta},
	\qquad
	\varepsilon_{\phi}{}^{\theta}=-\sin\theta.
\end{equation}
The metric perturbation may then be decomposed according to its tensorial character on the two-sphere. The $(t,r)$ components are expanded in scalar spherical harmonics as
\begin{equation}
	h_{ab}
	=
	\sum_{lm}h_{ab}^{lm}(t,r)\,Y_{lm}(\theta,\phi),
\end{equation}
the mixed components in vector spherical harmonics as
\begin{equation}
	h_{aB}
	=
	\sum_{lm}
	\left(
	j_{a}^{lm}(t,r)\,\psi_{B}^{lm}(\theta,\phi)
	+
	h_{a}^{lm}(t,r)\,\chi_{B}^{lm}(\theta,\phi)
	\right),
\end{equation}
and the purely angular components in tensor spherical harmonics as
\begin{equation}
	h_{AB}
	=
	\sum_{lm}
	\left(
	r^{2}K^{lm}(t,r)\,\bar g_{AB}Y_{lm}(\theta,\phi)
	+
	r^{2}G^{lm}(t,r)\,\psi_{AB}^{lm}(\theta,\phi)
	+
	h_{2}^{lm}(t,r)\,\chi_{AB}^{lm}(\theta,\phi)
	\right).
\end{equation}
In the Regge--Wheeler gauge, one sets
\begin{equation}
	j_{a}^{lm}=0,
	\qquad
	G^{lm}=0,
	\qquad
	h_{2}^{lm}=0,
\end{equation}
so that the surviving harmonic content of the metric perturbation takes a particularly simple form. The remaining nonvanishing perturbation amplitudes and their parity assignments are summarized in Table~\ref{tab:rw_gauge_decomposition}.
In particular, the angular sector may be chosen so that
\begin{equation}
	h_{\theta\phi}=0,
	\qquad
	h_{\phi\phi}=\sin^{2}\theta\,h_{\theta\theta}.
\end{equation}
With these choices, the spherical-harmonic decomposition removes the angular dependence from the first-order field equations, reducing the perturbation problem to a system of ordinary differential equations in the radial coordinate for each multipole $(l,m)$.

\begin{table}[t]
	\centering
	\renewcommand{\arraystretch}{1.2}
	\begin{tabular}{p{0.30\linewidth} p{0.52\linewidth} p{0.14\linewidth}}
		\hline
		\textbf{Metric sector} & \textbf{Harmonic expansion in RW gauge} & \textbf{Parity} \\
		\hline
		$h_{ab}$,\quad $a,b\in\{t,r\}$
		&
		$h_{ab}=\displaystyle\sum_{lm} h_{ab}^{lm}(t,r)\,Y_{lm}(\theta,\phi)$
		&
		even \\
		$h_{aB}$,\quad $a\in\{t,r\},\ B\in\{\theta,\phi\}$
		&
		$h_{aB}=\displaystyle\sum_{lm} h_{a}^{lm}(t,r)\,\chi_{B}^{lm}(\theta,\phi)$
		&
		odd \\
		$h_{AB}$,\quad $A,B\in\{\theta,\phi\}$
		&
		$h_{AB}=\displaystyle\sum_{lm} r^{2}K^{lm}(t,r)\,\bar g_{AB}\,Y_{lm}(\theta,\phi)$
		&
		even \\
		\hline
	\end{tabular}
	\caption{Regge--Wheeler gauge decomposition of static metric perturbations on a spherically symmetric background.}
	\label{tab:rw_gauge_decomposition}
\end{table}

\subsection{Exterior Solution of the Perturbed Configuration}
\label{subsec:exterior_perturbed}

We now consider the exterior region $r>R$, where the background geometry is the Schwarzschild spacetime \eqref{Schwarzschild_sol} with
$f(r)$ given by Eq.~\eqref{Schwarzschild_f}. Since $\bar T_{\mu\nu}=0$ in the exterior region, the background curvature satisfies
\begin{equation}
	\bar R_{\mu\nu}=0,
	\qquad
	\bar R=0.
\end{equation}
The linearized Einstein equations therefore reduce to
\begin{align}
	\delta R_{\mu\nu}-\frac{1}{2}\bar g_{\mu\nu}\,\delta R=0.
\end{align}
Taking the trace in four spacetime dimensions immediately gives $\delta R=0$, so that
\begin{align}
\label{vacuum_linearized_Rmunu}
	\delta R_{\mu\nu}=0,
	\qquad
	(r>R).
\end{align}
To simplify the analysis, we impose the conditions
\begin{equation}
\label{h_vac_constr}
	h_{tr}=h_{r\theta}=h_{r\phi}=0,
	\qquad
	h_{rr}=f^{-2}h_{tt}.
\end{equation}
Under these assumptions, the odd- and even-parity sectors decouple and may be treated separately.

\subsubsection{Odd-parity equations}

Setting $h_{tr}=0$, the $t\theta$ component of the linearized Ricci tensor becomes
\begin{align}
\label{odd_Rttheta_component_simplified}
	2\delta R_{t\theta}
	=
	\frac{1}{r^{2}\sin^{2}\theta}\partial_{\phi}\partial_{\theta}h_{t\phi}
	+
	\left(
	-f\partial_{r}^{2}
	-\frac{4M}{r^{3}}
	-\frac{1}{r^{2}\sin^{2}\theta}\partial_{\phi}^{2}
	\right)h_{t\theta}.
\end{align}
We now substitute the odd-parity harmonic expansions. Since the perturbations are static, the radial amplitudes depend only on $r$:
\begin{align}
	h_{t\theta}
	&=
	\sum_{lm}
	-h_{t}^{lm}(r)\frac{1}{\sin\theta}\partial_{\phi}Y_{lm}(\theta,\phi),
	\label{odd_htheta_expansion}\\
	h_{t\phi}
	&=
	\sum_{lm}
	h_{t}^{lm}(r)\sin\theta\,\partial_{\theta}Y_{lm}(\theta,\phi).
	\label{odd_htphi_expansion}
\end{align}
Carrying out the angular reduction and using the eigenvalue equation \eqref{scalar_sph_eigen}, one obtains
\begin{align}
\label{odd_Rttheta_harmonic_reduction}
	2\delta R_{t\theta}
	=
	\sum_{lm}
	\frac{1}{\sin\theta}
	\left(
	f\partial_{r}^{2}h_{t}^{lm}
	+\frac{4M}{r^{3}}h_{t}^{lm}
	-\frac{l(l+1)}{r^{2}}h_{t}^{lm}
	\right)
	\partial_{\phi}Y_{lm}(\theta,\phi),
\end{align}
where the intermediate steps are presented in Appendix~\ref{app:odd_parity_angular_reduction}. Therefore, the equation $\delta R_{t\theta}=0$ implies the radial equation
\begin{align}
\label{odd_parity_extr}
	r^{2}\partial_{r}^{2}h_{t}^{lm}
	+
	\frac{1}{f}
	\left(
	\frac{4M}{r}-l(l+1)
	\right)h_{t}^{lm}
	=
	0.
\end{align}
In the weak-field limit $M\to 0$, so that $f\to 1$, Eq.~\eqref{odd_parity_extr} reduces to
\begin{align}
	r^{2}\partial_{r}^{2}h_{t}^{lm}-l(l+1)h_{t}^{lm}=0,
\end{align}
which admits a growing solution proportional to $r^{l+1}$,
\begin{align}
	h_{t}^{lm}=k\,r^{l+1},
\end{align}
where $k$ is an integration constant. In the Schwarzschild exterior, we parameterize the solution as
\begin{align}
\label{odd_ht_solution_param}
	h_{t}^{lm}
	=
	\frac{2}{3(l-1)}\,b_{1}^{l}\,r^{l+1}B_{lm},
\end{align}
where $b_{1}^{l}\to 1$ as $r\to\infty$ and
\begin{align}
\label{odd_b1_def}
	b_{1}^{l}
	&=
	A_{3}^{l}
	-\frac{l+1}{l}\,2k_{l}^{mag}
	\left(\frac{R}{r}\right)^{2l+1}
	B_{3}^{l},\\
\label{odd_A3_def}
	A_{3}^{l}
	&=
	F\!\left(-l+1,\,-l-2;\,-2l;\,\frac{2M}{r}\right),\\
\label{odd_B3_def}
	B_{3}^{l}
	&=
	F\!\left(l-1,\,l+2;\,2l+2;\,\frac{2M}{r}\right).
\end{align}
Since there is no Newtonian analogue of the odd-parity sector, the coefficient $k_{l}^{mag}$ multiplying the $B_{3}^{l}$ branch in Eq.~\eqref{odd_b1_def} is identified as the magnetic-type gravitational Love number.

The remaining odd-parity equations act as consistency conditions. Starting from the explicit $t\phi$ component given in Appendix~\ref{app:odd_parity_angular_reduction}, one finds that $\delta R_{t\phi}=0$ reduces to the same radial equation \eqref{odd_parity_extr}. Finally, the $tr$ component can be written, using $h_{tr}=0$, as
\begin{align}
\label{odd_Rtr_component}
	2\delta R_{tr}
	=
	\frac{1}{r^{2}}
	\left(
	\partial_{\theta}\partial_{r}
	-2\partial_{r}\psi\,\partial_{\theta}
	+\cot\theta\,\partial_{r}
	-2\cot\theta\,\partial_{r}\psi
	\right)h_{t\theta}
	+\frac{1}{r^{2}\sin^{2}\theta}
	\left(
	\partial_{\phi}\partial_{r}
	-2\partial_{r}\psi\,\partial_{\phi}
	\right)h_{t\phi},
\end{align}
and substitution of Eqs.~\eqref{odd_htheta_expansion} and \eqref{odd_htphi_expansion} shows that $\delta R_{tr}=0$ is identically satisfied. Details are given in Appendix~\ref{app:odd_parity_tr_identity}.

\subsubsection{Even-parity equations}

We next turn to the even-parity sector of the static exterior perturbation. In the vacuum region $r>R$, the field equations reduce to
\begin{equation}
	\delta R_{\mu\nu}=0.
\end{equation}
In addition to the constraints \eqref{h_vac_constr}, we impose the angular identity
\begin{align}
\label{even_hphph_identity}
	h_{\phi\phi}=\sin^{2}\theta\,h_{\theta\theta}.
\end{align}
Starting from the general static expression for $\delta R_{tt}$, given in Appendix~\ref{app:linricci_diagonal}, and using the Schwarzschild relations
\begin{equation}
	e^{2\psi}=f,
	\qquad
	\psi'=\frac{f'}{2f},
\end{equation}
one finds
\begin{align}
\label{even_dRtt_general}
	2\delta R_{tt}
	=
	\left(
	-\partial_{r}^{2}f
	-\frac{2}{r}\partial_{r}f
	-f\partial_{r}^{2}
	-\frac{2f}{r}\partial_{r}
	-\frac{1}{r^{2}}D^{2}
	\right)h_{tt}
	+\frac{f}{r^{2}}\partial_{r}f
	\left(
	\partial_{r}-\frac{2}{r}
	\right)h_{\theta\theta}.
\end{align}
For the Schwarzschild function \eqref{Schwarzschild_f}, the first two terms vanish identically, and therefore one arrives at
\begin{align}
\label{even_dRtt_schwarzschild}
	2\delta R_{tt}
	=
	\left(
	-f\partial_{r}^{2}
	-\frac{2f}{r}\partial_{r}
	-\frac{1}{r^{2}}D^{2}
	\right)h_{tt}
	+\frac{f}{r^{2}}\partial_{r}f
	\left(
	\partial_{r}-\frac{2}{r}
	\right)h_{\theta\theta}.
\end{align}
We expand the even-parity amplitudes in scalar spherical harmonics:
\begin{align}
\label{even_harm_htt}
	h_{tt}(r,\theta,\phi)
	&=
	\sum_{lm}h_{tt}^{lm}(r)\,Y_{lm}(\theta,\phi),\\
\label{even_harm_hthth}
	h_{\theta\theta}(r,\theta,\phi)
	&=
	\sum_{lm}r^{2}K^{lm}(r)\,Y_{lm}(\theta,\phi).
\end{align}
Using Eq.~\eqref{scalar_sph_eigen}, one also finds
\begin{align}
\label{even_dhthth_identity}
	\left(
	\partial_{r}-\frac{2}{r}
	\right)h_{\theta\theta}
	=
	\sum_{lm}r^{2}\partial_{r}K^{lm}(r)\,Y_{lm}(\theta,\phi).
\end{align}
Substituting Eqs.~\eqref{even_harm_htt}--\eqref{even_dhthth_identity} into Eq.~\eqref{even_dRtt_schwarzschild} and using orthogonality yields, mode by mode,
\begin{align}
\label{even_dRtt_mode}
	\left(
	-f\partial_{r}^{2}
	-\frac{2f}{r}\partial_{r}
	+\frac{l(l+1)}{r^{2}}
	\right)h_{tt}^{lm}
	+
	f\,\partial_{r}f\,\partial_{r}K^{lm}
	=
	0.
\end{align}
For $l\geq 2$, we define
\begin{align}
\label{d_definition}
	d \equiv (l+2)(l-1).
\end{align}
To eliminate $K^{lm}$, we introduce the relation
\begin{align}
\label{even_K_relation}
	K^{lm}
	=
	\left(
	\frac{1}{f}
	+\frac{4M}{d r f}
	+\frac{2M}{d f}\partial_{r}
	\right)h_{tt}^{lm},
\end{align}
and differentiate with respect to $r$ to obtain
\begin{align}
\label{even_Kprime_relation}
	\partial_{r}K^{lm}
	=
	\left(
	-\frac{2Ml(l+1)}{d r^{2}f^{2}}
	+
	\left(
	\frac{1}{f}
	+\frac{4M(r-3M)}{d r^{2}f^{2}}
	\right)\partial_{r}
	+
	\frac{2M}{d f}\partial_{r}^{2}
	\right)h_{tt}^{lm}.
\end{align}
The derivation of Eq.~\eqref{even_Kprime_relation} is given in Appendix~\ref{app:evenparity_Kprime}. Substituting this expression into Eq.~\eqref{even_dRtt_mode}, using $\partial_{r}f=2M/r^{2}$, and simplifying, the $tt$ equation factorizes as
\begin{align}
\label{even_dRtt_factored}
	2\delta R_{tt}
	=
	\sum_{lm}
	\left(
	-\frac{1}{r^{2}}
	+\frac{4M^{2}}{d f r^{4}}
	\right)
	\left(
	fr^{2}\partial_{r}^{2}
	+2(r-3M)\partial_{r}
	-l(l+1)
	\right)h_{tt}^{lm}\,Y_{lm},
\end{align}
where the intermediate algebra is presented in Appendix~\ref{app:evenparity_master_derivation}. Hence $\delta R_{tt}=0$ implies the radial equation
\begin{align}
\label{even_master_equation}
	\left(
	fr^{2}\partial_{r}^{2}
	+2(r-3M)\partial_{r}
	-l(l+1)
	\right)h_{tt}^{lm}=0.
\end{align}
Equation \eqref{even_master_equation} may be used to eliminate second derivatives:
\begin{align}
\label{even_hpp_from_master}
	\partial_{r}^{2}h_{tt}^{lm}
	=
	\frac{l(l+1)}{fr^{2}}h_{tt}^{lm}
	-\frac{2(r-3M)}{fr^{2}}\partial_{r}h_{tt}^{lm}.
\end{align}
Substituting Eq.~\eqref{even_hpp_from_master} into Eq.~\eqref{even_Kprime_relation} gives the simpler identity
\begin{align}
\label{even_Kprime_simple}
	\partial_{r}K^{lm}
	=
	\frac{1}{f}\partial_{r}h_{tt}^{lm}.
\end{align}
Using the reduced component expressions summarized in Appendix~\ref{app:evenparity_consistency}, together with \eqref{h_vac_constr} and \eqref{even_Kprime_simple}, one finds that the remaining even-parity field equations are governed by the same master operator. In particular,
\begin{align}
\label{even_dRrr_master}
	2\delta R_{rr}
	&=
	-\frac{1}{f^{2}r^{2}}
	\left(
	fr^{2}\partial_{r}^{2}
	+2(r-3M)\partial_{r}
	-l(l+1)
	\right)h_{tt},\\
\label{even_dRthth_master}
	2\delta R_{\theta\theta}
	&=
	-\frac{1}{f}
	\left(
	fr^{2}\partial_{r}^{2}
	+2(r-3M)\partial_{r}
	-l(l+1)
	\right)h_{tt},\\
\label{even_dRphph_master}
	2\delta R_{\phi\phi}
	&=
	\sin^{2}\theta\,2\delta R_{\theta\theta}.
\end{align}
The off-diagonal components vanish identically under the same assumptions. For example,
\begin{align}
\label{even_dRrtheta_identity}
	2\delta R_{r\theta}
	&=
	\frac{1}{f}\partial_{r}\partial_{\theta}h_{tt}
	+
	\left(
	\frac{2}{r^{3}}\partial_{\theta}
	-\frac{1}{r^{2}}\partial_{r}\partial_{\theta}
	\right)h_{\theta\theta}
	=
	\sum_{lm}
	\left(
	\frac{1}{f}\partial_{r}h_{tt}^{lm}
	-\partial_{r}K^{lm}
	\right)\partial_{\theta}Y_{lm}
	=
	0,\\
\label{even_dRrphi_identity}
	2\delta R_{r\phi}
	&=
	\frac{1}{f}\partial_{r}\partial_{\phi}h_{tt}
	+
	\left(
	\frac{2}{r^{3}}\partial_{\phi}
	-\frac{1}{r^{2}}\partial_{r}\partial_{\phi}
	\right)h_{\theta\theta}
	=
	\sum_{lm}
	\left(
	\frac{1}{f}\partial_{r}h_{tt}^{lm}
	-\partial_{r}K^{lm}
	\right)\partial_{\phi}Y_{lm}
	=
	0,
\end{align}
where the last equalities follow from Eq.~\eqref{even_Kprime_simple}. Similarly, the $\theta\phi$ equation reduces to
\begin{align}
\label{even_dRthphi_identity}
	2\delta R_{\theta\phi}
	=
	\frac{1}{f}
	\left(
	\partial_{\theta}\partial_{\phi}
	-\cot\theta\,\partial_{\phi}
	\right)h_{tt}
	+
	f
	\left(
	-\partial_{\theta}\partial_{\phi}
	+\cot\theta\,\partial_{\phi}
	\right)h_{rr}
	=
	0,
\end{align}
after using $h_{rr}=f^{-2}h_{tt}$.

The solution of Eq.~\eqref{even_master_equation} must be linear in the tidal moments $\varepsilon_{lm}$. In the weak-field limit $M\to 0$, so that $f\to 1$, Eq.~\eqref{even_master_equation} admits the growing solution
\begin{equation}
	h_{tt}^{lm}=k\,r^{l}.
\end{equation}
For the Schwarzschild exterior, we therefore parameterize the asymptotically growing branch as
\begin{align}
\label{even_h_solution}
	h_{tt}^{lm}
	=
	-\frac{2}{l(l-1)}\,e_{1}^{l}(r)\,r^{l}\,\varepsilon_{lm},
\end{align}
where $e_{1}^{l}(r)\to 1$ as $r\to\infty$ and
\begin{align}
\label{e1_definition}
	e_{1}^{l}(r)
	=
	f^{2}A_{1}^{l}(r)
	+
	2k_{l}^{el}
	\left(\frac{R}{r}\right)^{2l+1}
	f^{2}B_{1}^{l}(r).
\end{align}
Here $A_{1}^{l}$ and $B_{1}^{l}$ are the hypergeometric functions
\begin{align}
\label{A1_definition}
	A_{1}^{l}(r)
	&=
	F\!\left(2-l,\,-l;\,-2l;\,\frac{2M}{r}\right),\\
\label{B1_definition}
	B_{1}^{l}(r)
	&=
	F\!\left(l+1,\,l+3;\,2l+2;\,\frac{2M}{r}\right),
\end{align}
and the coefficient $k_{l}^{el}$ multiplying the $B_{1}^{l}$ branch in Eq.~\eqref{e1_definition} is identified as the electric-type gravitational Love number. In the Newtonian limit $M/r\ll 1$, one has $f,A_{1}^{l},B_{1}^{l}\to 1$, and therefore
\begin{align}
\label{e1_newtonian_limit}
	e_{1}^{l}(r)
	=
	1
	+
	2k_{l}^{el}\left(\frac{R}{r}\right)^{2l+1}.
\end{align}
The corresponding expression for $K^{lm}$ is written as
\begin{align}
\label{even_K_solution}
	K^{lm}
	=
	-\frac{2}{l(l-1)}\,e_{2}^{l}(r)\,r^{l}\,\varepsilon_{lm},
\end{align}
where
\begin{align}
\label{e2_definition}
	e_{2}^{l}(r)
	=
	A_{2}^{l}(r)
	+
	2k_{l}^{el}
	\left(\frac{R}{r}\right)^{2l+1}
	B_{2}^{l}(r),
\end{align}
with
\begin{align}
\label{A2_definition}
	A_{2}^{l}(r)
	&=
	\frac{l+1}{l-1}
	F\!\left(-l,\,-l;\,-2l;\,\frac{2M}{r}\right)
	-\frac{2}{l-1}
	F\!\left(-l,\,-l-1;\,-2l;\,\frac{2M}{r}\right),\\
\label{B2_definition}
	B_{2}^{l}(r)
	&=
	\frac{l}{l+2}
	F\!\left(l+1,\,l+1;\,2l+2;\,\frac{2M}{r}\right)
	+\frac{2}{l+2}
	F\!\left(l+1,\,l;\,2l+2;\,\frac{2M}{r}\right).
\end{align}

\subsection{Interior Solution of the Perturbed Configuration}
\label{subsec:int_perturbed}

Inside the body, $r<R$, the matter sources the perturbation and the background scalar curvature is nonzero. The linearized Einstein equations therefore take the form
\begin{align}
\label{int_lin_einstein}
	\delta R_{\mu\nu}
	-\frac{1}{2}\bar g_{\mu\nu}\,\delta R
	-\frac{1}{2}h_{\mu\nu}\,\bar R
	=
	8\pi\,\delta T_{\mu\nu}.
\end{align}
Using Eq.~\eqref{scalar_curvature_interior}, this may be written more explicitly as
\begin{align}
\label{int_lin_einstein_explicit}
	\delta R_{\mu\nu}
	-\frac{1}{2}\bar g_{\mu\nu}\,\delta R
	-4\pi(\bar\rho-3\bar p)\,h_{\mu\nu}
	=
	8\pi\,\delta T_{\mu\nu},
\end{align}
where the perturbation of the stress-energy tensor is
\begin{equation}
\label{lin_Tmunu}
	\delta T_{\mu\nu}
	=
	(\bar\rho+\bar p)\left(\bar u_{\mu}\delta u_{\nu}+\bar u_{\nu}\delta u_{\mu}\right)
	+\bar u_{\mu}\bar u_{\nu}(\delta\rho+\delta p)
	+\bar g_{\mu\nu}\delta p
	+\bar p\,h_{\mu\nu}.
\end{equation}
Perturbing the background four-velocity \eqref{four_vels} to first order gives
\begin{align}
\label{pert_vels}
	\delta u^{\mu}
	=
	\frac{1}{2}e^{-3\psi}h_{tt}(1,0,0,0),
	\qquad
	\delta u_{\mu}
	=
	e^{-\psi}\left(\frac{1}{2}h_{tt},h_{tr},h_{t\theta},h_{t\phi}\right).
\end{align}
Substituting Eqs.~\eqref{four_vels} and \eqref{pert_vels} into Eq.~\eqref{lin_Tmunu} yields the nonvanishing components of $\delta T_{\mu\nu}$, which are listed in Appendix~\ref{app:int_deltaT_components}.

To describe the fluid deformation, we introduce the fractional radial displacement
\begin{align}
\label{int_F_def}
	F
	=
	\frac{\delta r}{r}
	=
	\sum_{lm}F^{lm}(r)\,Y_{lm}(\theta,\phi),
\end{align}
which gives rise to the Eulerian pressure and density perturbations
\begin{align}
\label{int_deltap}
	\delta p
	&=
	-r\bar p'\sum_{lm}F^{lm}(r)\,Y_{lm}(\theta,\phi),\\
\label{int_deltarho}
	\delta\rho
	&=
	-r\bar\rho'\sum_{lm}F^{lm}(r)\,Y_{lm}(\theta,\phi).
\end{align}
The unperturbed stellar structure is governed by the Tolman--Oppenheimer--Volkoff system. For the perturbation analysis that follows, the relevant background equations are
\begin{align}
\label{int_background_m_f}
	f(r)
	&=
	1-\frac{2m(r)}{r},
	\qquad
	m'(r)=4\pi r^{2}\bar\rho(r),\\
\label{int_background_psi_TOV}
	\psi'(r)
	&=
	\frac{m(r)+4\pi r^{3}\bar p(r)}{r^{2}f(r)},
	\qquad
	\bar p'(r)=-(\bar\rho+\bar p)\psi'(r).
\end{align}

As in the exterior analysis, the harmonic decomposition enforces
\begin{equation}
	h_{\theta\phi}=0,
	\qquad
	h_{\phi\phi}=\sin^{2}\theta\,h_{\theta\theta}.
\end{equation}
To simplify the interior problem further, we adopt the static constraints
\begin{align}
\label{int_constraints}
	h_{tr}=0,
	\qquad
	h_{r\theta}=0,
	\qquad
	h_{r\phi}=0,
	\qquad
	h_{rr}=f^{-1}e^{-2\psi}h_{tt}.
\end{align}
Under these assumptions, the odd- and even-parity sectors may again be analyzed separately.

\subsubsection{Odd-parity sector}

We begin with the axial (odd-parity) sector. The $t\theta$ component of Eq.~\eqref{int_lin_einstein_explicit}, together with $\bar g_{t\theta}=0$ and $\delta T_{t\theta}=-\bar\rho\,h_{t\theta}$, yields
\begin{align}
\label{int_odd_field_eq_ttheta}
	\delta R_{t\theta}+4\pi(3\bar p+\bar\rho)\,h_{t\theta}=0.
\end{align}
For the relevant components of $\delta T_{\mu\nu}$, see Appendix~\ref{app:int_deltaT_components}.
Using the coordinate expression for $\delta R_{t\theta}$ given in Appendix~\ref{app:int_deltaR_ttheta}, together with the odd-parity harmonic expansions
\begin{align}
	h_{t\theta}
	&=
	\sum_{lm}
	-h_{t}^{lm}(r)\,\frac{1}{\sin\theta}\,\partial_{\phi}Y_{lm}(\theta,\phi),
	\label{int_odd_h_ttheta}\\
	h_{t\phi}
	&=
	\sum_{lm}
	h_{t}^{lm}(r)\,\sin\theta\,\partial_{\theta}Y_{lm}(\theta,\phi),
	\label{int_odd_h_tphi}
\end{align}
and projecting onto $\partial_{\phi}Y_{lm}/\sin\theta$, 
while making use of the scalar-harmonic eigenvalue relation \eqref{scalar_sph_eigen}, one obtains
\begin{align}
\label{int_odd_intermediate}
	\Biggl(
	f\frac{d^{2}}{dr^{2}}
	+\left(\frac{f'}{2}-f\psi'\right)\frac{d}{dr}
	+\frac{4f}{r}\psi'
	-\frac{l(l+1)}{r^{2}}
	-8\pi(3\bar p+\bar\rho)
	\Biggr)h_{t}^{lm}(r)=0.
\end{align}
Eliminating $\psi'$ and $f'$ by means of Eqs.~\eqref{int_background_m_f} and \eqref{int_background_psi_TOV}, one arrives at the interior odd-parity master equation
\begin{align}
\label{int_odd_master}
	\Biggl(
	f\frac{d^{2}}{dr^{2}}
	-4\pi r(\bar\rho+\bar p)\frac{d}{dr}
	+\frac{4m}{r^{3}}
	-\frac{l(l+1)}{r^{2}}
	-8\pi(\bar\rho+\bar p)
	\Biggr)h_{t}^{lm}(r)=0,
\end{align}
where the intermediate algebra is presented in Appendix~\ref{app:int_deltaR_ttheta}.
Multiplying Eq.~\eqref{int_odd_master} by $r^{2}/f$ gives the more compact form
\begin{align}
\label{int_odd_compact}
	\left(
	r^{2}\frac{d^{2}}{dr^{2}}
	-r\,F(r)\frac{d}{dr}
	-G(r)
	\right)h_{t}^{lm}(r)=0,
\end{align}
where
\begin{align}
\label{int_odd_FG}
	F(r)
	&=
	\frac{4\pi r^{2}}{f}\left(\bar\rho+\bar p\right),
	\qquad
	G(r)
	=
	\frac{1}{f}
	\left(
	l(l+1)+8\pi r^{2}(\bar\rho+\bar p)-\frac{4m}{r}
	\right).
\end{align}
For numerical purposes, it is often convenient to evolve the logarithmic derivative
\begin{align}
\label{int_kl_def}
	k_{l}(r)
	:=
	r\,\frac{d h_{t}^{lm}/dr}{h_{t}^{lm}},
\end{align}
rather than the function $h_{t}^{lm}$ itself. In terms of $k_l(r)$, Eq.~\eqref{int_odd_compact} is equivalent to the first-order differential equation
\begin{align}
\label{int_kl_ode}
	r\frac{dk_{l}}{dr}
	+k_{l}(k_{l}-1)
	-F(r)k_{l}
	-G(r)
	=0,
\end{align}
which may be integrated once an equation of state is specified, so that $\bar\rho(r)$ and $\bar p(r)$ are known.

\subsubsection{Even-parity sector}

We now turn to the polar (even-parity) sector. The $tt$ component of Eq.~\eqref{int_lin_einstein_explicit} reads
\begin{align}
	\delta R_{tt}
	+\frac{1}{2}e^{2\psi}\delta R
	-4\pi(\bar\rho-3\bar p)h_{tt}
	=
	8\pi\left(e^{2\psi}\delta\rho-\bar\rho h_{tt}\right),
\end{align}
which may be rearranged as
\begin{align}
\label{int_even_tt_start}
	\delta R_{tt}
	+\frac{1}{2}e^{2\psi}\delta R
	+4\pi(3\bar p+\bar\rho)h_{tt}
	-8\pi e^{2\psi}\delta\rho
	=0.
\end{align}
To eliminate $\delta R$, we take the trace of Eq.~\eqref{int_lin_einstein} and use the background field equations, obtaining
\begin{align}
\label{int_deltaR_trace_id}
	\delta R
	=
	8\pi\left(
	h^{\mu\nu}\bar T_{\mu\nu}
	-\bar g^{\mu\nu}\delta T_{\mu\nu}
	\right).
\end{align}
Using the expressions for $\delta T_{\mu\nu}$ given in Appendix~\ref{app:int_deltaT_components}, together with the constraints \eqref{int_constraints} and the identity $h_{\phi\phi}=\sin^{2}\theta\,h_{\theta\theta}$, one finds
\begin{align}
	h^{\mu\nu}\bar T_{\mu\nu}
	&=
	e^{-2\psi}(\bar\rho+\bar p)\,h_{tt}
	+\frac{2\bar p}{r^{2}}\,h_{\theta\theta},
	\nonumber\\
	\bar g^{\mu\nu}\delta T_{\mu\nu}
	&=
	3\delta p-\delta\rho
	+e^{-2\psi}(\bar\rho+\bar p)\,h_{tt}
	+\frac{2\bar p}{r^{2}}\,h_{\theta\theta},
\end{align}
and therefore
\begin{align}
\label{int_deltaR_scalar}
	\delta R=8\pi(\delta\rho-3\delta p).
\end{align}
Substituting Eq.~\eqref{int_deltaR_scalar} into Eq.~\eqref{int_even_tt_start} yields the reduced $tt$ equation
\begin{align}
\label{int_even_tt_reduced}
	\delta R_{tt}
	+4\pi(3\bar p+\bar\rho)\,h_{tt}
	-4\pi e^{2\psi}\left(3\delta p+\delta\rho\right)
	=0.
\end{align}
Using the component expression for $\delta R_{tt}$ given in Appendix~\ref{app:int_deltaR_tt}, together with the constraint \eqref{int_constraints}, the harmonic identity \eqref{scalar_sph_eigen}, and the first-order relation
\begin{align}
\label{int_Kprime_relation}
	\frac{d}{dr}K^{lm}
	=
	e^{-2\psi}\frac{d}{dr}h_{tt}^{lm},
\end{align}
which we adopt following Ref.~\cite{yang2022tidal}, one obtains
\begin{align}
\label{int_deltaR_tt_simplified}
	2\delta R_{tt}^{lm}
	=
	\Biggl(
	-f\frac{d^{2}}{dr^{2}}
	+\left(
	4\pi r(3\bar p+\bar\rho)-\frac{2(r-3m)}{r^{2}}
	\right)\frac{d}{dr}
	-8\pi(3\bar p+\bar\rho)
	+\frac{l(l+1)}{r^{2}}
	\Biggr)h_{tt}^{lm}.
\end{align}
For further details, see Appendix~\ref{app:int_deltaR_tt}. Substituting Eq.~\eqref{int_deltaR_tt_simplified} into Eq.~\eqref{int_even_tt_reduced} and multiplying through by $-2r^{2}/f$ gives
\begin{align}
\label{int_even_intermediate}
	\Biggl(
	r^{2}\frac{d^{2}}{dr^{2}}
	+\left(
	\frac{2(r-3m)}{f}
	-\frac{4\pi r^{3}(3\bar p+\bar\rho)}{f}
	\right)\frac{d}{dr}
	-\frac{l(l+1)}{f}
	\Biggr)h_{tt}^{lm}
	+\frac{8\pi r^{2}}{f}e^{2\psi}\left(3\delta p+\delta\rho\right)^{lm}
	=0.
\end{align}

To close the system, we relate the displacement $F^{lm}$ to the metric perturbation by means of the perturbed hydrostatic equilibrium condition
\begin{align}
\label{int_hydr_perturbed}
	\partial_{\mu}\delta p
	+(\bar\rho+\bar p)\delta a_{\mu}
	+(\delta\rho+\delta p)\bar a_{\mu}
	=0,
\end{align}
where $a_{\mu}=u^{\sigma}\nabla_{\sigma}u_{\mu}$ and $\bar a_{\mu}=(0,\psi',0,0)$. In the static limit, the perturbed acceleration is
\begin{align}
\label{int_delta_a}
	\delta a_{\mu}
	=
	\frac{1}{2}e^{-2\psi}
	\left(
	0,\,
	2\psi' h_{tt}-\partial_{r}h_{tt},\,
	-\partial_{\theta}h_{tt},\,
	-\partial_{\phi}h_{tt}
	\right),
\end{align}
with details given in Appendix~\ref{app:int_accel}.
Choosing $\mu=\theta$ in Eq.~\eqref{int_hydr_perturbed}, so that $\bar a_{\theta}=0$, substituting Eq.~\eqref{int_deltap} and $\delta a_{\theta}=-(1/2)e^{-2\psi}\partial_{\theta}h_{tt}$ from Eq.~\eqref{int_delta_a}, and projecting onto $\partial_{\theta}Y_{lm}$, one finds
\begin{align}
	F^{lm}
	=
	-\frac{\bar\rho+\bar p}{2r\bar p'}\,e^{-2\psi}h_{tt}^{lm}.
\end{align}
Using the TOV relation $\bar p'=-(\bar\rho+\bar p)\psi'$ from Eq.~\eqref{int_background_psi_TOV}, this becomes
\begin{align}
\label{int_F_lm}
	F^{lm}
	=
	\frac{e^{-2\psi}}{2r\psi'}\,h_{tt}^{lm}.
\end{align}
With Eqs.~\eqref{int_deltap}, \eqref{int_deltarho}, and \eqref{int_F_lm}, the source term in Eq.~\eqref{int_even_intermediate} becomes
\begin{align}
\label{int_even_source}
	\frac{8\pi r^{2}}{f}e^{2\psi}\left(3\delta p+\delta\rho\right)
	&=
	-\frac{8\pi r^{3}}{f}e^{2\psi}\left(3\bar p'+\bar\rho'\right)F
	\nonumber\\
	&=
	\frac{4\pi r^{2}}{f}\left(\bar\rho+\bar p\right)
	\left(
	3+\frac{d\bar\rho}{d\bar p}
	\right)h_{tt},
\end{align}
where we have used $\bar\rho'=(d\bar\rho/d\bar p)\bar p'$ together with $\bar p'=-(\bar\rho+\bar p)\psi'$. Substituting Eq.~\eqref{int_even_source} into Eq.~\eqref{int_even_intermediate}, we obtain the closed second-order equation
\begin{align}
\label{int_even_master}
	\Biggl(
	r^{2}\frac{d^{2}}{dr^{2}}
	+\left(
	\frac{2(r-3m)}{f}
	-\frac{4\pi r^{3}(3\bar p+\bar\rho)}{f}
	\right)\frac{d}{dr}
	-\frac{l(l+1)}{f}
	+\frac{4\pi r^{2}(\bar\rho+\bar p)}{f}
	\left(
	3+\frac{d\bar\rho}{d\bar p}
	\right)
	\Biggr)h_{tt}^{lm}
	=0.
\end{align}
It is useful to define
\begin{align}
\label{int_A_def}
	A(r)
	&\equiv
	\frac{2}{f}
	\left(
	1-\frac{3m(r)}{r}-2\pi r^{2}(3\bar p+\bar\rho)
	\right),\\
\label{int_B_def}
	B(r)
	&\equiv
	\frac{1}{f}
	\left(
	l(l+1)
	-4\pi r^{2}(\bar\rho+\bar p)
	\left(
	3+\frac{d\bar\rho}{d\bar p}
	\right)
	\right),
\end{align}
in terms of which Eq.~\eqref{int_even_master} takes the compact form
\begin{align}
\label{int_even_master_compact}
	r^{2}\frac{d^{2}h_{tt}^{lm}}{dr^{2}}
	+rA(r)\frac{dh_{tt}^{lm}}{dr}
	-B(r)h_{tt}^{lm}
	=0.
\end{align}
Finally, introducing the logarithmic derivative
\begin{align}
\label{int_eta_def}
	\eta_{l}(r)
	:=
	\frac{r\,d h_{tt}^{lm}/dr}{h_{tt}^{lm}},
\end{align}
one may rewrite Eq.~\eqref{int_even_master_compact} as the first-order equation
\begin{align}
\label{int_eta_ode}
	r\frac{d\eta_{l}}{dr}
	+\eta_{l}(\eta_{l}-1)
	+A(r)\eta_{l}
	-B(r)
	=0,
\end{align}
supplemented by the regular-center boundary condition
\begin{align}
	\eta_{l}(0)=l.
\end{align}
In the next section, we specialize the general interior even-parity system to the quadrupolar case and show explicitly how it reduces to the standard master equation used in the relativistic Love-number literature.

\section{From the General Perturbation Setup to Hinderer's Equation}
\label{sec:hinderer_reduction}

In Sec.~\ref{sec:perturbed_configuration}, we derived the static interior even-parity perturbation equations for a relativistic perfect-fluid star in a general $(l,m)$ decomposition.
The corrected center expansion derived in the next section is stated in the standard quadrupolar variable $H(r)$, so we first need an explicit bridge from the general interior system to that form. This section provides that bridge: specializing to the quadrupolar sector $l=2$, we reduce the general equations to the single master equation used by Hinderer~\cite{Hinderer2008}, fixing the conventions needed for the new center analysis.

Throughout this section, a prime denotes differentiation with respect to the radial coordinate $r$. The unperturbed spacetime of a static, spherically symmetric star is written in Schwarzschild-like coordinates as
\begin{align}
\label{background_metric}
	ds^2
	=
	-e^{\nu(r)}dt^2
	+
	e^{\lambda(r)}dr^2
	+
	r^2\left(d\theta^2+\sin^2\theta\,d\phi^2\right),
\end{align}
where $\nu(r)$ and $\lambda(r)$ are the background metric potentials. We now perturb the metric according to
\begin{equation}
\label{metric_pert_def}
	g_{\alpha\beta}
	=
	\bar g_{\alpha\beta}+h_{\alpha\beta},
	\qquad
	h_{\alpha\beta}=\delta g_{\alpha\beta},
\end{equation}
and restrict attention to the static, even-parity, quadrupolar sector in Regge-Wheeler gauge. In this gauge, the perturbation is diagonal and may be written as
\begin{equation}
\label{RW_ansatz}
	h_{\alpha\beta}
	=
	\mathrm{diag}
	\!\left(
	-e^{\nu}H_{0}(r),\,
	e^{\lambda}H_{2}(r),\,
	r^{2}K(r),\,
	r^{2}\sin^{2}\theta\,K(r)
	\right)
	Y_{2m}(\theta,\phi).
\end{equation}
Thus, the entire quadrupolar even-parity perturbation is encoded in the three radial amplitudes $H_{0}(r)$, $H_{2}(r)$, and $K(r)$. In the unperturbed configuration, the fluid is at rest in these coordinates, so the background four-velocity $\bar u^{\alpha}$ has only a temporal component. Imposing the normalization condition to first order then yields
\begin{align}
\label{delta_u_rels}
	\delta u^{\alpha}
	&=
	\frac{1}{2}e^{-3\nu/2}h_{00}\,\delta^{\alpha}{}_{0},
	\qquad
	\delta u_{\alpha}
	=
	\frac{1}{2}e^{-\nu/2}h_{00}\,\delta^{0}{}_{\alpha}.
\end{align}
For the linearized Einstein equations, it is useful to work with the mixed components of the perturbed stress-energy tensor, $\delta T^{\alpha}{}_{\beta}$. Using the relation $\delta g^{\alpha\gamma}=-h^{\alpha\gamma}$, we obtain
\begin{align}
\label{delta_T_mixed_def}
	\delta T^{\alpha}{}_{\beta}
	=
	\bar g^{\alpha\gamma}\delta T_{\gamma\beta}
	+
	\delta g^{\alpha\gamma}T_{\gamma\beta}.
\end{align}
Varying Eq.~\eqref{perfect_fluid_Tmunu} and substituting Eq.~\eqref{delta_u_rels}, together with the background relations, gives the compact identities
\begin{align}
\label{deltaT_diag_pressure}
	\delta T^{t}{}_{t}
	=
	-\delta\rho,
	\qquad
	\delta T^{r}{}_{r}
	=
	\delta T^{\theta}{}_{\theta}
	=
	\delta T^{\phi}{}_{\phi}
	=
	\delta p,
\end{align}
while all off-diagonal components vanish. We also impose the local adiabatic relation
\begin{equation}
\label{adiabatic_rel}
	\delta\rho
	=
	\frac{d\rho}{dp}\,\delta p.
\end{equation}
The linearized Einstein equations take the form
\begin{align}
\label{lin_Einstein}
	\delta G^{\alpha}{}_{\beta}
	=
	8\pi\,\delta T^{\alpha}{}_{\beta}.
\end{align}
To evaluate $\delta G^{\alpha}{}_{\beta}$ explicitly, we recall the standard first-order geometric identities. The perturbation of the Christoffel symbols is
\begin{align}
\label{pert_Chr}
	\delta\Gamma^{\alpha}{}_{\beta\gamma}
	&=
	\frac{1}{2}
	\left(
	-h_{\beta\gamma}{}^{;\alpha}
	+h^{\alpha}{}_{\gamma;\beta}
	+h^{\alpha}{}_{\beta;\gamma}
	\right),
\end{align}
which may equivalently be written as
\begin{align}
\label{deltaGamma_partial}
	\delta\Gamma^{\alpha}{}_{\beta\gamma}
	&=
	\frac{1}{2}\bar g^{\alpha\delta}
	\left(
	\partial_{\beta}h_{\delta\gamma}
	+
	\partial_{\gamma}h_{\delta\beta}
	-
	\partial_{\delta}h_{\beta\gamma}
	\right)
	-
	\bar g^{\alpha\delta}\Gamma^{\epsilon}{}_{\beta\gamma}h_{\delta\epsilon}.
\end{align}
Varying the Ricci tensor gives then
\begin{equation}
\label{deltaRicci_expanded}
	\delta R_{\alpha\beta}
	=
	\partial_{\rho}\delta\Gamma^{\rho}{}_{\beta\alpha}
    -
	\partial_{\beta}\delta\Gamma^{\rho}{}_{\rho\alpha}
	+
	\Gamma^{\rho}{}_{\rho\lambda}\delta\Gamma^{\lambda}{}_{\beta\alpha}
	+
	\delta\Gamma^{\rho}{}_{\rho\lambda}\Gamma^{\lambda}{}_{\beta\alpha}
	-
    \Gamma^{\rho}{}_{\beta\lambda}\delta\Gamma^{\lambda}{}_{\rho\alpha}
	-
	\delta\Gamma^{\rho}{}_{\beta\lambda}\Gamma^{\lambda}{}_{\rho\alpha},
\end{equation}
while the perturbed Ricci scalar is
\begin{align}
\label{delta_R_scalar}
	\delta R
	=
	\bar g^{\alpha\beta}\delta R_{\alpha\beta}
	-
	h^{\alpha\beta}R_{\alpha\beta}.
\end{align}
Finally, the mixed components of the linearized Einstein tensor are
\begin{align}
\label{pert_mixed_G}
	\delta G^{\alpha}{}_{\beta}
	=
	\delta R^{\alpha}{}_{\beta}
	-
	\frac{1}{2}\delta^{\alpha}{}_{\beta}\,\delta R.
\end{align}
We first consider the difference of the angular components of the field equations. Using Eqs.~\eqref{deltaT_diag_pressure} and \eqref{pert_mixed_G}, one finds
\begin{align}
\label{pert_G_the_phi_diff}
	\delta G^{\theta}{}_{\theta}
	-
	\delta G^{\phi}{}_{\phi}
	&=
	\frac{H_{0}+H_{2}}{r^{2}}
	\left(
	\frac{\cot\theta}{2}\partial_{\theta}Y_{2m}
	-
	\frac{1}{2}\partial_{\theta}^{2}Y_{2m}
	+
	\frac{1}{2\sin^{2}\theta}\partial_{\phi}^{2}Y_{2m}
	\right)
	=
	0.
\end{align}
Hence, the radial amplitudes are not independent but must satisfy
\begin{equation}
	H_{0}=-H_{2}.
\end{equation}
It is, therefore, natural to introduce a single master function,
\begin{equation}
\label{rel_H}
	H(r)\equiv H_{0}(r)=-H_{2}(r).
\end{equation}
Next, since $\delta T^{r}{}_{\theta}=0$, the $(r,\theta)$ component of Eq.~\eqref{lin_Einstein} requires $\delta G^{r}{}_{\theta}=0$. Evaluating this component for the Regge–Wheeler ansatz \eqref{RW_ansatz} and using Eq.~\eqref{rel_H}, one obtains
\begin{align}
	\delta G^{r}{}_{\theta}
	=
	-\frac{e^{-\lambda}}{2}
	\left(
	H'+K'+\nu'H
	\right)
	\partial_{\theta}Y_{2m}(\theta,\phi)
	=
	0,
\end{align}
which implies the first-order constraint
\begin{align}
\label{K_prime_eq}
	K'(r)
	=
	-H'(r)-\nu'(r)H(r).
\end{align}
Once Eqs.~\eqref{rel_H} and \eqref{K_prime_eq} are imposed, the angular field equations simplify considerably. In particular,
\begin{align}
	\delta G^{\theta}{}_{\theta}
	=
	\delta G^{\phi}{}_{\phi}
	=
	-\frac{e^{-\lambda}}{2r}
	\left(
	\lambda'+\nu'
	\right)
	H\,Y_{2m}(\theta,\phi)
	=
	8\pi\,\delta p,
\end{align}
and hence, the pressure perturbation is
\begin{align}
\label{pert_p}
	\delta p
	=
	-\frac{e^{-\lambda}}{16\pi r}
	\left(
	\lambda'+\nu'
	\right)
	H\,Y_{2m}(\theta,\phi).
\end{align}
We now form the combination $(t,t)-(r,r)$ of the linearized Einstein equations. On the matter side, using Eqs.~\eqref{deltaT_diag_pressure}, \eqref{adiabatic_rel}, and \eqref{pert_p}, we obtain
\begin{align}
\label{deltaGtt_rr_matter}
	\delta G^{t}{}_{t}
	-
	\delta G^{r}{}_{r}
	=
	\left(
	1+\frac{d\rho}{dp}
	\right)
	\frac{e^{-\lambda}}{2r}
	\left(
	\lambda'+\nu'
	\right)
	H\,Y_{2m}(\theta,\phi).
\end{align}
On the geometric side, substituting Eq.~\eqref{RW_ansatz}, imposing Eqs.~\eqref{rel_H} and \eqref{K_prime_eq}, and using the $l=2$ eigenvalue equation \eqref{scalar_sph_eigen}, one finds
\begin{equation}
\label{deltaGtt_rr_geom}
	\delta G^{t}{}_{t}
	-
	\delta G^{r}{}_{r}
	=
	e^{-\lambda}
	\Biggl(
	-H''
	+
	\left(
	\frac{\lambda'}{2}
	-
	\frac{\nu'}{2}
	-
	\frac{2}{r}
	\right)H'
	+
	\left(
	\frac{\lambda'\nu'}{2}
	+
	\frac{(\nu')^{2}}{2}
	-
	\nu''
	-
	\frac{\lambda'}{r}
	-
	\frac{3\nu'}{r}
	+
	\frac{6e^{\lambda}}{r^{2}}
	\right)H
	\Biggr)
	Y_{2m}(\theta,\phi).
\end{equation}
Equating Eqs.~\eqref{deltaGtt_rr_matter} and \eqref{deltaGtt_rr_geom}, and dividing by the common factor $e^{-\lambda}Y_{2m}$, yields
\begin{align}
\label{H_master_intermediate}
	H''
	+
	\left(
	\frac{\nu'}{2}
	-
	\frac{\lambda'}{2}
	+
	\frac{2}{r}
	\right)H'
	+
	\Biggl(
	\nu''
	-
	\frac{\lambda'\nu'}{2}
	-
	\frac{(\nu')^{2}}{2}
	+
	\frac{3\lambda'}{2r}
	+
	\frac{7\nu'}{2r}
	-
	\frac{6e^{\lambda}}{r^{2}}
	+
	\frac{\lambda'+\nu'}{2r}\frac{d\rho}{dp}
	\Biggr)H
	=
	0.
\end{align}
To bring this equation into the standard Tolman--Oppenheimer--Volkoff form, we now rewrite the coefficients in terms of the background stellar variables. The metric function $\lambda(r)$ is related to the enclosed mass function $m(r)$ by
\begin{equation}
\label{tov_eq_1}
	e^{-\lambda}
	=
	1-\frac{2m}{r},
\end{equation}
while the mass profile satisfies
\begin{equation}
\label{tov_eq_2}
	\frac{dm}{dr}
	=
	4\pi r^{2}\rho.
\end{equation}
Differentiating Eq.~\eqref{tov_eq_1} and using Eq.~\eqref{tov_eq_2} gives
\begin{equation}
\label{tov_eq_3}
	\lambda'
	=
	e^{\lambda}
	\left(
	8\pi r\rho-\frac{2m}{r^{2}}
	\right).
\end{equation}
The second metric potential obeys
\begin{equation}
\label{tov_eq_4}
	\nu'
	=
	e^{\lambda}
	\left(
	8\pi rp+\frac{2m}{r^{2}}
	\right),
\end{equation}
and hydrostatic equilibrium is encoded in the TOV equation
\begin{equation}
\label{tov_eq_5}
	\frac{dp}{dr}
	=
	-\frac{\rho+p}{2}\,\nu'.
\end{equation}
From these relations one readily derives the useful identities
\begin{align}
\label{nu_lambda_rels}
	\frac{\nu'+\lambda'}{r}
	&=
	8\pi e^{\lambda}(\rho+p),\\
	\frac{\nu'-\lambda'}{2}
	&=
	e^{\lambda}
	\left(
	\frac{2m}{r^{2}}+4\pi r(p-\rho)
	\right),
\end{align}
and
\begin{align}
\label{nu_second_der}
	\nu''
	=
	\frac{1}{2}\nu'\lambda'
	-
	\frac{1}{2}(\nu')^{2}
	+
	\frac{\nu'+\lambda'}{r}
	+
	\frac{2}{r^{2}}\left(1-e^{\lambda}\right).
\end{align}
Substituting Eqs.~\eqref{nu_lambda_rels} and \eqref{nu_second_der} into Eq.~\eqref{H_master_intermediate}, and simplifying, we arrive at the interior static even-parity master equation in the form quoted by Hinderer:
\begin{align}
\label{2nd_ord_eq_fin}
	H''
	+\mathcal{W}(r)\,H'
	+\mathcal{Z}(r)\,H
	=
	0,
\end{align}
where
\begin{align}
\label{coeff_W}
	\mathcal{W}(r)
	&=
	\frac{2}{r}
	+
	e^{\lambda}
	\left(
	\frac{2m}{r^{2}}+4\pi r(p-\rho)
	\right),\\
\label{coeff_Z}
	\mathcal{Z}(r)
	&=
	-\frac{6e^{\lambda}}{r^{2}}
	+
	4\pi e^{\lambda}
	\left(
	5\rho+9p+(\rho+p)\frac{d\rho}{dp}
	\right)
	-
	(\nu')^{2}.
\end{align}
This equation is precisely the quadrupolar interior master equation used in the standard relativistic Love-number analysis. The derivation above makes clear that it is not an independent starting point, but rather the specialized $l=2$ reduction of the general interior even-parity perturbation system derived in the previous section.


\section{Center Expansion for $H(r)$: Correcting Hinderer's Eq.~(16)}
\label{sec:center_expansion}

We now turn to the regular behavior of the interior master function near the stellar center. This point is important both analytically and numerically. Analytically, regularity at $r=0$ determines the admissible local solution of the tidal equation. Numerically, the near-center expansion provides the initial data from which the integration of Eq.~\eqref{2nd_ord_eq_fin} is started. Since Hinderer's original treatment quoted the regular expansion in compact form, it is worthwhile to derive it explicitly and carefully verify the subleading coefficient.

We therefore study the $l=2$ equation in Regge--Wheeler gauge directly near $r=0$, using a local Frobenius expansion together with the TOV equations. To make the equation-of-state dependence explicit, we introduce the adiabatic index
\begin{align}
\label{adb_indx}
	\Gamma(\rho)
	=
	\frac{d\ln p}{d\ln\rho}
	=
	\frac{\rho}{p}\,\frac{dp}{d\rho}.
\end{align}
Expanding the equation of state about the central density $\rho_{c}$ gives
\begin{align}
\label{eos_exps}
	p(\rho)
	&=
	p_{c}
	+
	\frac{\Gamma_{c}p_{c}}{\rho_{c}}(\rho-\rho_{c})
	+
	\frac{p_{c}}{2\rho_{c}^{2}}
	\Biggl[
	\Gamma_{c}^{2}
	-\Gamma_{c}
	+
	\Gamma_{c}
	\left(
	\frac{d\ln\Gamma}{d\ln\rho}
	\right)_{c}
	\Biggr]
	(\rho-\rho_{c})^{2}
	+
	\mathcal{O}\!\left((\rho-\rho_{c})^{3}\right).
\end{align}
Similarly, we Taylor-expand the inverse derivative $d\rho/dp$ about $p_{c}$:
\begin{equation}
\label{drho_dp_eq}
	\frac{d\rho}{dp}
	=
	\left.\frac{d\rho}{dp}\right|_{p=p_{c}}
	+
	\left.\frac{d^{2}\rho}{dp^{2}}\right|_{p=p_{c}}(p-p_{c})
	+
	\mathcal{O}\!\left((p-p_{c})^{2}\right).
\end{equation}
Near the center, we seek analytic solutions of the TOV system and therefore adopt the series ansätze
\begin{align}
\label{series_ansatz}
	m(r)
	&=
	\sum_{k\geq0}m_{k}r^{k},
	\qquad
	p(r)
	=
	\sum_{k\geq0}p_{k}r^{k},
	\qquad
	\rho(r)
	=
	\sum_{k\geq0}\rho_{k}r^{k}.
\end{align}
Substituting Eq.~\eqref{series_ansatz} into Eqs.~\eqref{tov_eq_2}--\eqref{tov_eq_5} and matching the first nonvanishing orders yields
\begin{align}
\label{m_series}
	m(r)
	&=
	\frac{4\pi}{3}\rho_{c}r^{3}
	-
	\frac{8\pi^{2}}{15}
	\frac{\rho_{c}}{\Gamma_{c}p_{c}}
	(\rho_{c}+p_{c})(\rho_{c}+3p_{c})\,r^{5}
	+
	\mathcal{O}(r^{7}),\\[4pt]
	p(r)
	&=
	p_{c}
	-
	\frac{2\pi}{3}(\rho_{c}+p_{c})(\rho_{c}+3p_{c})\,r^{2}
	+
	\mathcal{O}(r^{4}),\\[4pt]
\label{rho_series}
	\rho(r)
	&=
	\rho_{c}
	-
	\frac{\rho_{c}}{\Gamma_{c}p_{c}}
	\frac{2\pi}{3}
	(\rho_{c}+p_{c})(\rho_{c}+3p_{c})\,r^{2}
	+
	\mathcal{O}(r^{4}).
\end{align}
Using Eq.~\eqref{tov_eq_1}, we then obtain
\begin{equation}
\label{elambda_series}
	e^{\lambda}
	=
	1+\frac{8\pi}{3}\rho_{c}r^{2}
	+
	\mathcal{O}(r^{4}).
\end{equation}
Integrating Eq.~\eqref{tov_eq_4} gives
\begin{equation}
\label{nu_series}
	\nu(r)
	=
	\nu_{0}
	+
	4\pi
	\left(
	p_{c}+\frac{\rho_{c}}{3}
	\right)r^{2}
	-
	4\pi^{2}
	\left(
	p_{c}^{2}
	-\frac{\rho_{c}^{2}}{9}
	+\frac{p_{c}\rho_{c}}{5\Gamma_{c}}
	+\frac{4\rho_{c}^{2}}{15\Gamma_{c}}
	+\frac{\rho_{c}^{3}}{15\Gamma_{c}p_{c}}
	\right)r^{4}
	+
	\mathcal{O}(r^{6}),
\end{equation}
so that
\begin{equation}
\label{nuprime_series}
	\nu'(r)
	=
	8\pi
	\left(
	p_{c}+\frac{\rho_{c}}{3}
	\right)r
	-
	16\pi^{2}
	\left(
	p_{c}^{2}
	-\frac{\rho_{c}^{2}}{9}
	+\frac{p_{c}\rho_{c}}{5\Gamma_{c}}
	+\frac{4\rho_{c}^{2}}{15\Gamma_{c}}
	+\frac{\rho_{c}^{3}}{15\Gamma_{c}p_{c}}
	\right)r^{3}
	+
	\mathcal{O}(r^{5}).
\end{equation}
Combining Eq.~\eqref{eos_exps} with Eq.~\eqref{rho_series} allows us to re-express Eq.~\eqref{drho_dp_eq} as a radial expansion:
\begin{align}
\label{eq_d_rho_d_p}
	\frac{d\rho}{dp}(r)
	&=
	\frac{\rho_{c}}{\Gamma_{c}p_{c}}
	+
	\frac{2\pi\rho_{c}(\rho_{c}+p_{c})(\rho_{c}+3p_{c})}{3p_{c}^{2}\Gamma_{c}^{3}}
	\Biggl(
	\Gamma_{c}^{2}
	-
	\Gamma_{c}
	+
	\Gamma_{c}
	\left(
	\frac{d\ln\Gamma}{d\ln\rho}
	\right)_{c}
	\Biggr)r^{2}
	+
	\mathcal{O}(r^{4}).
\end{align}
We now return to the master equation \eqref{2nd_ord_eq_fin}, which we write in the compact form
\begin{align}
\label{mast_eq_concs}
	H''
	+
	\mathcal{A}(r)H'
	+
	\mathcal{B}(r)H
	=
	0.
\end{align}
Using the TOV expansions above, the coefficient functions admit the near-center expansions
\begin{align}
\label{A_series}
	\mathcal{A}(r)
	&=
	\frac{2}{r}
	+
	\left(
	4\pi p_{c}-\frac{4\pi\rho_{c}}{3}
	\right)r
	+
	\mathcal{A}_{3}r^{3}
	+
	\mathcal{O}(r^{5}),\\
\label{B_series}
	\mathcal{B}(r)
	&=
	-\frac{6}{r^{2}}
	+
	\left(
	36\pi p_{c}
	+
	4\pi\rho_{c}
	+
	\frac{4\pi\rho_{c}}{\Gamma_{c}}
	+
	\frac{4\pi\rho_{c}^{2}}{\Gamma_{c}p_{c}}
	\right)
	+
	\mathcal{B}_{2}r^{2}
	+
	\mathcal{O}(r^{4}).
\end{align}
Since Eq.~\eqref{mast_eq_concs} has a regular singular point at $r=0$, we seek a Frobenius solution of the form
\begin{align}
\label{H_frob}
	H(r)
	=
	r^{s}\sum_{n=0}^{\infty}a_{n}r^{n},
	\qquad
	a_{0}\neq0.
\end{align}
Substituting Eqs.~\eqref{A_series}--\eqref{H_frob} into Eq.~\eqref{mast_eq_concs} yields
\begin{align}
\label{diff_pwr_series}
	&\sum_{n=0}^{\infty}
	\Bigl[(n+s)(n+s+1)-6\Bigr]a_{n}r^{n+s-2}
	+
	\sum_{n=0}^{\infty}
	\Bigl[\alpha_{0}(n+s)+C_{0}\Bigr]a_{n}r^{n+s}
	\nonumber\\
	&\qquad
	+
	\sum_{n=0}^{\infty}
	\Bigl[\mathcal{A}_{3}(n+s)+\mathcal{B}_{2}\Bigr]a_{n}r^{n+s+2}
	+\cdots
	=
	0,
\end{align}
where we have introduced the shorthand
\begin{align}
	\alpha_{0}
	&=
	4\pi
	\left(
	p_{c}-\frac{\rho_{c}}{3}
	\right),
	\qquad
	C_{0}
	=
	4\pi\rho_{c}
	\left(
	1+\frac{1}{\Gamma_{c}}+\frac{\rho_{c}}{\Gamma_{c}p_{c}}
	\right)
	+
	36\pi p_{c}.
\end{align}
The lowest-order term, proportional to $r^{s-2}$, yields the indicial equation
\begin{align}
	s^{2}+s-6=0,
\end{align}
whose roots are
\begin{align}
	s_{1}=2,
	\qquad
	s_{2}=-3.
\end{align}
Regularity at the origin selects the larger root, $s=2$. The recurrence relation then becomes
\begin{equation}
	\Bigl[(k+2)(k+3)-6\Bigr]a_{k}
	+
	\Bigl[\alpha_{0}k+C_{0}\Bigr]a_{k-2}
	+
	\Bigl[\mathcal{A}_{3}(k-2)+\mathcal{B}_{2}\Bigr]a_{k-4}
	=
	0,
	\qquad
	k\geq0,
\end{equation}
with the convention that coefficients with negative indices vanish. Solving successively, one finds
\begin{align}
	a_{1}
	&=
	a_{3}
	=
	0,
	\qquad
	a_{2}
	=
	-\frac{2\alpha_{0}+C_{0}}{14}\,a_{0},
	\nonumber\\
	a_{4}
	&=
	\frac{(4\alpha_{0}+C_{0})(2\alpha_{0}+C_{0})-14(2\mathcal{A}_{3}+\mathcal{B}_{2})}{504}\,a_{0}.
\end{align}
Hence, the regular Frobenius expansion of the master function is
\begin{align*}
	H(r)
	&=
	a_{0}r^{2}
	\Biggl(
	1
	-
	\frac{2\alpha_{0}+C_{0}}{14}\,r^{2}
	+
	\frac{(4\alpha_{0}+C_{0})(2\alpha_{0}+C_{0})-14(2\mathcal{A}_{3}+\mathcal{B}_{2})}{504}\,r^{4}
	+
	\mathcal{O}(r^{6})
	\Biggr)
	\\
	&=
	a_{0}r^{2}
	\Biggl(
	1
	-
	\frac{2\pi}{7}
	\left(
	11p_{c}
	+\frac{\rho_{c}}{3}
	+\frac{\rho_{c}}{\Gamma_{c}}
	+\frac{\rho_{c}^{2}}{\Gamma_{c}p_{c}}
	\right)r^{2}
	+\cdots
	\Biggr).
\end{align*}
Equivalently, the near-center expansion may be written in the compact form
\begin{align}
\label{H_near_center}
	H(r)
	=
	a_{0}r^{2}
	\Biggl(
	1
	-
	\frac{2\pi}{7}
	\Biggl(
	11p_{c}
	+\frac{\rho_{c}}{3}
	+
	(p_{c}+\rho_{c})
	\left(
	\frac{d\rho}{dp}
	\right)_{c}
	\Biggr)r^{2}
	+\cdots
	\Biggr).
\end{align}
Equation~\eqref{H_near_center} is the regular center expansion that should be used when initializing the numerical integration of the tidal master equation. The leading behavior $H\sim r^{2}$ is fixed by regularity, while the subleading coefficient carries the first nontrivial dependence on the central thermodynamic data. This is the point at which the correction to the commonly quoted expression enters.

\section{Exterior Solution}
\label{sec:exterior_solution}

In the exterior region, $r>R$, the spacetime is a vacuum. Accordingly, the matter variables vanish, and the enclosed mass becomes constant:
\begin{align}
	\rho=0, \qquad p=0, \qquad m(r)=M.
\end{align}
The exterior background is therefore the Schwarzschild geometry. In this region, the interior master equation \eqref{2nd_ord_eq_fin} simplifies considerably. Introducing the dimensionless variable
\begin{align}
	\xi=\frac{r}{M}-1,
\end{align}
the $l=2$ equation reduces to
\begin{align}
	(\xi^{2}-1)\frac{d^{2}H}{d\xi^{2}}
	+2\xi\frac{dH}{d\xi}
	-\left(6+\frac{4}{\xi^{2}-1}\right)H=0.
\end{align}
This is the associated Legendre equation with degree $2$ and order $2$, and its general solution is
\begin{align}
\label{H_extr_in_xi}
	H(\xi)
	=
	c_{1}\,Q_{2}^{2}(\xi)
	+
	c_{2}\,P_{2}^{2}(\xi),
\end{align}
where $P_{2}^{2}(\xi)$ and $Q_{2}^{2}(\xi)$ are the associated Legendre functions of the first and second kind, respectively, and $c_{1}$ and $c_{2}$ are integration constants to be fixed by matching to the stellar interior. To extract the physical meaning of these two branches, we return to the original radial coordinate and examine the large-$r$ behavior of Eq.~\eqref{H_extr_in_xi}. Using the explicit forms of the associated Legendre functions, one finds
\begin{align}
\label{H_asym}
	H(r)
	=
	\frac{8}{5}\,c_{1}\,\frac{M^{3}}{r^{3}}
	+
	3\,c_{2}\,\frac{r^{2}}{M^{2}}
	+\mathcal{O}\!\left(\frac{r}{M}\right)
	+\mathcal{O}\!\left(\frac{M^{4}}{r^{4}}\right).
\end{align}
The two independent asymptotic behaviors have a clear physical interpretation: the growing branch is associated with the externally applied tidal field, while the decaying branch encodes the induced quadrupolar response of the star. To make this identification precise, we compare the asymptotic exterior solution with the far-field expansion of the metric component $g_{tt}$. We write~\cite{ThorneHartle1985, Thorne1998} 
\begin{align}
\label{g_tt_asym}
	\frac{1-g_{tt}}{2}
	&=
	-\frac{M}{r}
	-\frac{3Q_{ij}}{2r^{3}}
	\left(
	n^{i}n^{j}-\frac{1}{3}\delta^{ij}
	\right)
	+\mathcal{O}\!\left(\frac{1}{r^{3}}\right)
	+\frac{1}{2}\mathcal{E}_{ij}x^{i}x^{j}
	+\mathcal{O}\!\left(r^{3}\right),
\end{align}
where $n^{i}=x^{i}/r$ is the unit radial vector, $\mathcal{E}_{ij}$ is the external quadrupolar tidal field, and $Q_{ij}$ is the induced quadrupole moment of the star. In the linear-response regime, these quantities are related by
\begin{align}
\label{linear_tidal_rel}
	Q_{ij}
	=
	-\lambda_{t}\,\mathcal{E}_{ij}.
\end{align}
The corresponding dimensionless quadrupolar Love number is defined by
\begin{align}
\label{def_k2}
	k_{2}
	=
	\frac{3}{2}\lambda_{t}R^{-5}.
\end{align}
The constants in the exterior solution are determined by matching the asymptotic form \eqref{H_asym} to the quadrupolar part of the far-field metric \eqref{g_tt_asym}. Using Eq.~\eqref{linear_tidal_rel} to eliminate the induced quadrupole moment in favor of the tidal deformability, one finds, for a single $l=2$ harmonic mode,
\begin{align}
\label{coeffs}
	c_{1}
	=
	\frac{15}{8}\frac{\lambda_{t}\mathcal{E}}{M^{3}},
	\qquad
	c_{2}
	=
	\frac{1}{3}M^{2}\mathcal{E},
\end{align}
where $\mathcal{E}$ denotes the amplitude of the chosen quadrupolar tidal mode.

The final step is to match the exterior solution to the regular interior solution at the stellar surface $r=R$. Evaluating both $H(r)$ and $H'(r)$ at the surface, and eliminating the constants $c_{1}$, $c_{2}$, and the tidal amplitude $\mathcal{E}$ in favor of the surface data, one obtains the Love number in closed form:
\begin{align}
\label{def_Love_num}
	k_2
	&=
	\frac{8C^5}{5}(1-2C)^2
	\left[2+2C(y_s-1)-y_s\right]
	\times \nonumber\\
	&\quad
	\Biggl\{
	2C\left[6-3y_s+3C(5y_s-8)\right]
	+4C^3\left[13-11y_s+C(3y_s-2)+2C^2(1+y_s)\right]
	\nonumber\\
	&\qquad\qquad
	+3(1-2C)^2
	\left[
	2-y_s+2C(y_s-1)
	\right]
	\log(1-2C)
	\Biggr\}^{-1}.
\end{align}
Here,
\begin{align}
	y_s
	=
	\left.\frac{rH'(r)}{H(r)}\right|_{r=R}
\end{align}
is the surface value of the logarithmic derivative of the even-parity master function. Equation~\eqref{def_Love_num} provides the desired link between the interior perturbation problem and the observable tidal response. Once the interior equation is integrated from the center to the surface and $y_s$ is determined, the quadrupolar Love number $k_2$ follows immediately from the compactness $C$ and the matching formula above.

\section{Numerical Results}
\label{sec:numerical_results}

In this section, we outline the numerical procedure used to compute  
the quadrupolar Love numbers and to quantify the impact of the corrected regular-center expansion of $H(r)$. We adopt a polytropic equation of state in the form~\cite{Tooper1964,Tooper1965}
\begin{align}
	p = \kappa \rho^\Gamma \,,
\end{align}
where the adiabatic index is $\Gamma = 1 + 1/n$, $\kappa$ is a constant, and $n$ is the polytropic index. Throughout the numerical calculations reported here, we set $\kappa=1$ as a convenient normalization. For each prescribed pair $(n,C_{\rm tar})$, where $C_{\rm tar}$ is one of the target compactnesses listed in Table~\ref{k_2_comp}, the central density $\rho_c$ is determined by a one-dimensional shooting procedure. A trial value of $\rho_c$ fixes the central pressure through $p_c=\kappa \rho_c^\Gamma$, after which the stellar-structure equations are integrated outward until the pressure falls to zero within numerical tolerance, thereby defining the stellar surface. The corresponding radius $R(\rho_c)$ and enclosed mass $M(\rho_c)$ determine the model compactness $C_{\rm model}=M(\rho_c)/R(\rho_c)$. We then solve
\begin{align}
	\mathcal{R}(\rho_c) = C_{\rm model}-C_{\rm tar} = 0 \,.
\end{align}
If $\mathcal{R}(\rho_c)>0$, the trial density is decreased, whereas if $\mathcal{R}(\rho_c)<0$, it is increased. An automatic bracketing step is first used to identify an interval in $\rho_c$ over which $\mathcal{R}$ changes sign, and the matched value of $\rho_c$ is then obtained with Brent’s method~\cite{brent2013algorithms}. The iteration is terminated once the relative compactness mismatch satisfies $|C_{\rm model}-C_{\rm tar}|/C_{\rm tar}<10^{-10}$.

To compute the Love numbers, we integrate the interior tidal master equation~\eqref{2nd_ord_eq_fin} simultaneously with the TOV background equations. Specifically, the background configuration is obtained from Eqs.~\eqref{tov_eq_2}--\eqref{tov_eq_5}, while the tidal equation \eqref{2nd_ord_eq_fin} is rewritten as a first-order system by introducing $\Phi(r) \equiv H'(r)$, so that one solves
\begin{align}
	H' = \Phi \,, \qquad  
	\Phi' = -\mathcal{W}(r)\,\Phi -\mathcal{Z}(r)\,H \,,
\end{align}
where $\mathcal{W}(r)$ and $\mathcal{Z}(r)$ are the coefficient functions defined in Eqs.~\eqref{coeff_W} and \eqref{coeff_Z}. The integration is started at a small but finite radius $r_0=10^{-6}$. The regular center expansions for $m(r)$ and $p(r)$ supply the initial background data, while the perturbation is initialized as
\begin{equation}
\label{num_center_H_generic}	
	H(r_0)=a_0 r_0^2\bigl(1 -\sigma r_0^2\bigr) \,, \qquad
	H'(r_0)=2a_0 r_0-4a_0 \,\sigma r_0^3 \,.
\end{equation}
To isolate the effect of the corrected near-center behavior, we perform two otherwise identical integrations that differ only in the choice of $\sigma$. In the first run, we use Hinderer's original coefficient,
\begin{equation}
\label{num_alpha_H}	
	\sigma_{\rm H} =\frac{2\pi}{7}\left(5\rho_c+9p_c
	+(\rho_c+p_c)\left(\frac{d\rho}{dp}\right)_{c}
	\right),
\end{equation}
whereas in the second run we use the corrected coefficient implied by Eq.~\eqref{H_near_center},
\begin{equation}
\label{num_alpha_corr}	
	\sigma_{\rm corr} =\frac{2\pi}{7}\left( 11p_c+\frac{\rho_c}{3}
	+(\rho_c+p_c)\left(\frac{d\rho}{dp}\right)_{c}
	\right).
\end{equation}
Since $k_2$ depends on the ratio $H'(R)/H(R)$, the overall normalization is irrelevant, and we may therefore set $a_0=1$ without loss of generality.

At the stellar surface, we evaluate $y_s$ and then compute the quadrupolar Love numbers from Eq.~\eqref{def_Love_num}. The results are summarized in Table~\ref{k_2_comp}. In the table, $k_{2}^{(\mathrm{H})}$ denotes the Love numbers obtained when the perturbation is initialized with $\sigma_{\rm H}$, whereas $k_{2}^{(\mathrm{corr})}$ 
denotes those obtained with the corrected coefficient $\sigma_{\rm corr}$. The table reports results for polytropic indices $n\in\{0.3,\,0.5,\,0.7,\,1.0,\,1.2\}$ and compactnesses ranging from the near-Newtonian regime ($C\sim 10^{-5}$) to strongly relativistic configurations ($C=0.25$). The expected qualitative trends are clearly reproduced: at fixed $n$, the Love number decreases monotonically with increasing compactness, reflecting the weakening of the tidal response by stronger relativistic self-gravity; at fixed compactness, $k_2$ likewise decreases as $n$ increases, consistent with the greater central concentration of softer stellar models.
\begin{table}[htbp]
	\centering
	\small
	\renewcommand{\arraystretch}{1.2}
	\begin{tabular}{|c|c|c|c|}
		\hline
		$n$ & $C$ & $k_{2}^{(\mathrm{H})}$ & $k_{2}^{(\mathrm{corr})}$ \\
		\hline
		0.3 & $10^{-5}$ & 0.551016 & 0.551016 \\
		\hline
		0.3 & 0.10 & 0.287361 & 0.287361 \\
		\hline
		0.3 & 0.15 & 0.195545 & 0.195545 \\
		\hline
		0.3 & 0.20 & 0.126031 & 0.126031 \\
		\hline
		0.5 & $10^{-5}$ & 0.449103 & 0.449103 \\
		\hline
		0.5 & 0.10 & 0.228420 & 0.228420 \\
		\hline
		0.5 & 0.15 & 0.153139 & 0.153139 \\
		\hline
		0.5 & 0.20 & 0.096760 & 0.096760 \\
		\hline
		0.5 & 0.25 & 0.056184 & 0.056184 \\
		\hline
		0.7 & $10^{-5}$ & 0.362622 & 0.362622 \\
		\hline
		0.7 & 0.10 & 0.179921 & 0.179921 \\
		\hline
		0.7 & 0.15 & 0.118491 & 0.118491 \\
		\hline
		0.7 & 0.20 & 0.073039 & 0.073039 \\
		\hline
		0.7 & 0.25 & 0.040830 & 0.040830 \\
		\hline
		1.0 & $10^{-5}$ & 0.259892 & 0.259892 \\
		\hline
		1.0 & 0.10 & 0.123207 & 0.123207 \\
		\hline
		1.0 & 0.15 & 0.078371 & 0.078371 \\
		\hline
		1.0 & 0.20 & 0.045886 & 0.045886 \\
		\hline
		1.0 & 0.25 & 0.023454 & 0.023454 \\
		\hline
		1.2 & $10^{-5}$ & 0.206178 & 0.206178 \\
		\hline
		1.2 & 0.10 & 0.094109 & 0.094109 \\
		\hline
		1.2 & 0.15 & 0.058033 & 0.058033 \\
		\hline
		1.2 & 0.20 & 0.032303 & 0.032303 \\
		\hline
	\end{tabular}
	\caption{Comparison of the quadrupolar Love numbers $k_{2}^{(\mathrm{H})}$ and $k_{2}^{(\mathrm{corr})}$, obtained with the near-center coefficients \eqref{num_alpha_H} and \eqref{num_alpha_corr}. Here $k_{2}^{(\mathrm{H})}$ denotes the Love number introduced by Hinderer and $k_{2}^{(\mathrm{corr})}$ denotes the Love number obtained via the equation \eqref{num_alpha_corr}.}
\label{k_2_comp}	
\end{table}

The key result of Table~\ref{k_2_comp} is that the corrected center expansion does not alter the final Love numbers for the parameter range studied here. Defining
$\Delta\sigma \equiv \sigma_{\mathrm corr}-\sigma_{\mathrm H}$,
the two near-center prescriptions differ by
\begin{equation}
	\Delta H(r_0) = -a_0 \Delta\sigma \, r_0^4 \,, \qquad
	\Delta H'(r_0) = -4a_0 \Delta\sigma \, r_0^3 \,.
\end{equation}
Hence, the correction modifies only the subleading part of the regular solution, while the physically dominant leading behavior $H\sim a_0 r^2$ is identical in both cases. Because the integration starts at the small radius $r_0=10^{-6}$, the factors $r_0^4$ and $r_0^3$ suppress the difference so strongly that it remains numerically invisible in the surface quantity $y_s$, and therefore in $k_2$, for all models considered.

We therefore conclude that the expansion \eqref{H_near_center} is the analytically correct regular-center solution and should be used in the formal development of the perturbation problem. However, for the present numerical setup and for the range of $(n,C)$ explored here, replacing Hinderer’s original coefficient with the corrected one has no practical effect on the computed Love numbers.


\section{Static Tidal Perturbations of SdS: Master Equation}
\label{sec:sds_master}

This section derives the static, even-parity tidal master equation on a Schwarzschild–de Sitter background by reducing the linearized vacuum equations to a single radial ODE for the relevant metric amplitude. We start with a four-dimensional, static and spherically symmetric background geometry written in Schwarzschild-like coordinates
\begin{align}
	\label{metric_init}
	ds^{2}= -f(r)\,dt^{2} +f^{-1}(r)\,dr^{2} +r^{2}\left(d\theta^{2}+\sin^{2}\theta\,d\phi^{2}\right) ,
\end{align}
where the metric function is given by
\begin{align}
	\label{eq_f_r}	
	f(r) = e^{2\psi(r)} = 1-\frac{2M}{r}-\frac{\Lambda r^{2}}{3} \,.
\end{align}
Here, $M$ denotes the black-hole mass and $\Lambda$ is the cosmological constant. In what follows, we focus on $\Lambda>0$, so the background is the Schwarzschild--de Sitter (SdS) solution. The field equations are the vacuum Einstein equations with cosmological constant
\begin{align}
	R_{\mu\nu}-\frac{1}{2}\,g_{\mu\nu}R+\Lambda g_{\mu\nu}= 0 \,.
\end{align}
We now perturb the metric according to
\begin{align}
	g_{\mu\nu} = \bar{g}_{\mu\nu} + h_{\mu\nu}\,,
	\qquad \delta g_{\mu\nu} = h_{\mu\nu}\,.
\end{align}
Since the SdS background is an Einstein manifold, $R_{\mu\nu}=\Lambda g_{\mu\nu}$, the linearized vacuum equations can be written in the form
\begin{align}
	\delta R_{\mu\nu}= \Lambda h_{\mu\nu}\,.
\end{align}
We restrict to static, even-parity perturbations and work in Regge--Wheeler gauge, imposing
$h_{t r} = h_{r\theta} = h_{r\phi} = 0$. To further fix the remaining gauge freedom, we additionally enforce the diagonal relation $h_{rr}=f^{-2}(r)\,h_{tt}$. 
With these choices, the $(r,\theta)$ component of the linearized equations simplifies dramatically: it reduces to $\delta R_{r\theta}=0$, which implies
\begin{align}
	\label{eq_var_Ricci}
	\partial_{\theta}\!\Bigl(
	\frac{1}{2f}\,\partial_{r}h_{tt}
	+\Bigl(\frac{1}{r^{3}}-\frac{1}{2r^{2}}\partial_{r}\Bigr)h_{\theta\theta}\Bigr) = 0 \,.
\end{align}
To separate variables, we expand the angular dependence in scalar spherical harmonics,
\begin{align}
	h_{\theta\theta}(r,\theta,\phi)= \sum_{l, m} r^{2}K^{l m}(r)\,
	Y_{l m}(\theta,\phi) ,
\end{align}
and likewise
\begin{align}
	h_{tt}(r,\theta,\phi)=\sum_{l, m} 
	h_{tt}^{l m}(r)\,Y_{l m}(\theta,\phi) .
\end{align}
Substituting these decompositions into Eq.~\eqref{eq_var_Ricci} and projecting onto a given $(l,m)$ mode using the orthogonality of $Y_{l m}$ yields the radial constraint
\begin{align}
	\label{radl_constr}
	\partial_r K^{l m}=\frac{1}{f}\,\partial_r h_{tt}^{l m}\,.
\end{align}
Using the quantity $d$ defined in Eq.~\eqref{d_definition}, we employ the remaining independent components of the linearized equations to eliminate $K^{lm}$ in favor of $h_{tt}^{lm}$.
One parametrization that achieves this efficiently is
\begin{align}
	\label{def_K}
	K^{l m}=
	\left(\frac{1}{f}
	+\frac{4}{d\,r\,f}\Biggl(M+\frac{\Lambda r^{3}}{6}\Biggr)
	+\frac{2}{d\,f}\Biggl(M+\frac{\Lambda r^{3}}{6}\Biggr)\partial_{r}
	\right)h_{tt}^{l m} \,.
\end{align}
Inserting Eq.~\eqref{def_K} into Eq.~\eqref{radl_constr}
and simplifying leads to a second-order ODE for $h_{tt}^{l m}(r)$. Since the background is spherically symmetric, the resulting equation depends on $l$, but not on $m$. Henceforth, we suppress the $(l,m)$ label and write $h(r) = h_{tt}^{l m}(r)$.
The resulting master equation takes the form:
\begin{equation}
	\label{master_eq}	
	f(r)\,r^{2}h''(r)
	+\Biggl(\frac{3\Lambda r^{4}f(r)}{6M+\Lambda r^{3}}
	+2(r-3M) \Biggr)h'(r)
	 +\Biggl(
	\frac{6\Lambda r^{3}f(r)}{6M+\Lambda r^{3}} +2\Lambda r^{2}
	+\frac{3d\,\Lambda r^{3}}{6M+\Lambda r^{3}} -l(l+1)
	\Biggr)h(r) = 0 \,.
\end{equation}
In the asymptotically flat limit $\Lambda=0$, the SdS background reduces to the Schwarzschild geometry, and the master equation admits a closed-form solution for each static, even-parity multipole $l\ge 2$. The general exterior solution can be written in terms of associated Legendre functions (namely $P_l^{2}$ and $Q_l^{2}$); see Appendix~\ref{app:Lambda0} for a compact derivation and the explicit expression.

\subsection{Master Equation in Horizon Variables}

We now specialize to $\Lambda>0$ (SdS) and introduce the
de Sitter radius $L$ via $L^{2}= 3/\Lambda$, so that
\begin{align}
	f(r)=1-\frac{2M}{r}-\frac{r^{2}}{L^{2}} \,.
\end{align}
The horizon radii are the real roots of $f(r)=0$. In the two-horizon regime $(0<3\sqrt{3}\,M/L<1)$, there are two
distinct positive roots, denoted by $r_b$ (black-hole horizon) and $r_c$ (cosmological horizon), together with a third (negative) root $r_n$. One explicit parametrization is
\begin{align}
	r_b &= \frac{2L}{\sqrt{3}}\,
	\sin\!\left(
	\frac{1}{3}\arcsin\!\left(\frac{3\sqrt{3}\,M}{L}\right)
	\right) , \label{rb_explicit}\\
	r_c &= \frac{2L}{\sqrt{3}}\,
	\sin\!\left(
	\frac{1}{3}\arcsin\!\left(\frac{3\sqrt{3}\,M}{L}\right)
	+\frac{2\pi}{3}
	\right) , \label{rc_explicit}\\
	r_n&=-(r_b+r_c)\,,
	\label{rn_def}
\end{align}
with the ordering $r_c>r_b>0>r_n$. In terms of these roots, the metric function factorizes as
\begin{align}
	\label{f_factorized}	
	f(r)=\frac{(r-r_b)(r_c-r)(r-r_n)}{rL^{2}}\,.
\end{align}
The horizon radii also satisfy the useful identities
\begin{subequations}
	\label{root_idents}
	\begin{align}
		\label{root_identities_a}	
		r_b^{2}+r_b r_c+r_c^{2} &= L^{2} \,,  \\
		r_b r_c(r_b+r_c) &= 2L^{2}M \,. 
		\label{root_identities_b}
	\end{align}
\end{subequations}
Dividing Eq.~\eqref{master_eq} by $f r^{2}$ and using Eq.~\eqref{f_factorized} together with Eqs.~\eqref{root_idents}, one may trade the parameters $(M,L)$ for $(r_b,r_c)$. The resulting master equation in horizon variables reads
\begin{align}
	\label{diff_eq_hrzn}
	h''(r)
	&+\Bigg( \frac{3 r^{2}}{r^{3}+r_{b}r_{c}(r_{b}+r_{c})}
	+\frac{2\bigl(r_{b}^{2}+r_{b}r_{c}+r_{c}^{2}\bigr)}
	{(r-r_{b})(r_{c}-r)\bigl(r+r_{b}+r_{c}\bigr)} 
	-\frac{3\,r_{b}r_{c}(r_{b}+r_{c})}
	{r\,(r-r_{b})(r_{c}-r)\bigl(r+r_{b}+r_{c}\bigr)}
	\Bigg)h'(r) \nonumber \\[6pt]
	&\quad
	+\frac{l(l+1)\bigl(r_{b}^{2}+r_{b}r_{c}+r_{c}^{2}\bigr)
		\Bigl(2r^{3}-r_{b}r_{c}(r_{b}+r_{c})\Bigr)}
	{r\,(r-r_{b})(r_{c}-r)\bigl(r+r_{b}+r_{c}\bigr)
		\bigl(r^{3}+r_{b}r_{c}(r_{b}+r_{c})\bigr)} h(r) = 0.
\end{align}
 We study this equation in the near-Nariai regime \cite{Nariai, Nariai2} in our forthcoming paper \cite{Altas}.


\section{Conclusions}
\label{sec:conclusions}
In this work, we presented a detailed analysis of static tidal perturbations and gravitational Love numbers for relativistic, spherically symmetric stars in general relativity. Working in the Regge--Wheeler gauge and using a decomposition in scalar, vector, and tensor spherical harmonics, we derived the odd- and even-parity perturbation equations in both the vacuum exterior and the perfect-fluid interior. In the even-parity sector, we showed explicitly how the general perturbation system reduces to the familiar quadrupolar master equation used in relativistic tidal-deformability calculations.

Our main analytic result is a corrected derivation of the regular-center expansion of the interior even-parity master function. A careful Frobenius analysis shows that the subleading coefficient of the regular solution differs from the expression commonly quoted in the literature. This correction is important at the formal level since it clarifies the precise local structure of the interior solution and fixes the correct near-center data for the perturbation problem.

We then investigated the numerical impact of this corrected center expansion for polytropic equations of state. For the stellar models considered in this work, we found that the final quadrupolar Love number $k_{2}$ is unchanged within numerical accuracy. This outcome is physically reasonable: the correction enters only at subleading powers in the regular initial data, while the dominant leading behavior near the center remains unchanged. Thus, the corrected expansion is analytically necessary, even though its practical effect on $k_{2}$ is negligible for the parameter range studied here.

A further result of this paper is the extension of the static even-parity analysis to the Schwarzschild–de Sitter background. In that setting, the usual asymptotically flat exterior problem is replaced by a two-horizon geometry, and the standard interpretation of tidal response must accordingly be reconsidered. We derived the corresponding master equation in both the usual radial variables and a horizon-based parameterization, thereby providing a natural starting point for future studies of tidal perturbations in non-asymptotically flat spacetimes.

Taken together, these results provide a self-contained account of relativistic static tidal perturbation theory while sharpening several technical points that are often treated only briefly: the full reduction to the quadrupolar master equation, the correct regular-center structure of the interior solution, and the extension of the formalism to a background with a cosmological horizon. We hope that this presentation will be useful both for practical computations of relativistic Love numbers and for further investigations of tidal response in more general gravitational settings.

\begin{acknowledgments}
	We thank T. Hinderer for the correspondence about the center expansion (\ref{H_near_center}). E. A., E. K., and O. O. are supported by the TUBITAK Grant No. 123F353. E.K. acknowledges the support of the Outstanding Young Scientist Award of
The Turkish Academy of Sciences (TUBA-GEBIP).

\end{acknowledgments}

\bibliographystyle{unsrt}
\bibliography{refs}


\appendix

\section{Background geometry: connection and curvature components}
\label{app:background_geometry}

This appendix summarizes the Levi-Civita connection and curvature for the background metric \eqref{bg_metric}. 

\subsection{Nonzero Christoffel symbols}

For the line element \eqref{bg_metric}, the non-vanishing connection coefficients are:
\begin{align}
	\bGamma^t{}_{tr} &= \psi' \,,
	&
	\bGamma^r{}_{rr} &= -\frac{f'}{2f} \,,
	&
	\bGamma^r{}_{tt} &= f\e^{2\psi}\psi' \,,
	\label{Gamma_set1} \\
	\bGamma^\theta{}_{r\theta} &= \frac{1}{r} \,,
	&
	\bGamma^\phi{}_{r\phi} &= \frac{1}{r} \,,
	&
	\bGamma^r{}_{\theta\theta} &= -rf \,,
	\label{Gamma_set2} \\
	\bGamma^r{}_{\phi\phi} &= -rf\sin^2\theta \,,
	&
	\bGamma^\theta{}_{\phi\phi} &= -\sin\theta\cos\theta \,,
	&
	\bGamma^\phi{}_{\theta\phi} &= \cot\theta \,.
	\label{Gamma_set3}
\end{align}

\subsection{Ricci tensor and scalar}

The independent nonzero Ricci components can be written as
\begin{align}
	\bR_{tt} &= \e^{2\psi}f\left(\psi''
    +(\psi')^2
    +\frac{\psi'f'}{2f}+\frac{2\psi'}{r}\right),
	\label{Rtt_bg} \\
	\bR_{rr} &= -\psi''-(\psi')^2-\frac{\psi'f'}{2f}-\frac{f'}{rf} \,,
	\label{Rrr_bg} \\
	\bR_{\theta\theta} &= f\left(-\frac{r f'}{2f}-r\psi'-1\right)+1 \,,
	\qquad
	\bR_{\phi\phi}=\sin^2\theta\,\bR_{\theta\theta} \,.
	\label{eq:Rthth_bg}
\end{align}
The scalar curvature $\bR=\bg^{\mu\nu}\bR_{\mu\nu}$ reduces to
\begin{align}
	\bR = -2f(\psi')^{2}-2f\psi''-f'\psi'-\frac{4f}{r}\psi'
		-\frac{2f'}{r}-\frac{2f}{r^{2}}+\frac{2}{r^{2}} \,.
\end{align}

\section{Linearized Ricci tensor in the static case}
\label{app:linricci_static}

This appendix collects the coordinate components of the linearized Ricci tensor $\delta R_{\mu\nu}$ for static perturbations of the background metric
\eqref{bg_metric}.

\subsection{Diagonal components}
\label{app:linricci_diagonal}

\subsubsection{$tt$ component}

\begin{align}
	2\delta R_{tt} = {}&
	f^{2}e^{2\psi}\Bigl(
	-\psi'\partial_{r}
	-\frac{2f'}{f}\psi'
	-2(\psi')^{2}
	-2\psi''
	-\frac{4}{r}\psi'
	\Bigr)h_{rr}
	\nonumber\\
	&+\left(
	-f\partial_{r}^{2}
	+\left(2f\psi'-\frac{f'}{2}-\frac{2f}{r}\right)\partial_{r}
	-2f(\psi')^{2}
	-\frac{1}{r^{2}}D^{2}
	\right)h_{tt}
	+\frac{2fe^{2\psi}}{r^{2}}\psi'\left(-\partial_{\theta}-\cot\theta\right)h_{r\theta}
	\nonumber\\
	&-\frac{2fe^{2\psi}}{r^{2}\sin^{2}\theta}\psi'\,\partial_{\phi}h_{r\phi}+\frac{fe^{2\psi}}{r^{2}}\psi'\left(\partial_{r}-\frac{2}{r}\right)h_{\theta\theta}
	+\frac{fe^{2\psi}}{r^{2}\sin^{2}\theta}\psi'\left(\partial_{r}-\frac{2}{r}\right)h_{\phi\phi}.
	\label{Rtt_explicit}
\end{align}

\subsubsection{$rr$ component}

\begin{align}
	2\delta R_{rr} = {}&
	\left(
	\left(f\psi'+\frac{2f}{r}\right)\partial_{r}
	+\psi' f'
	+\frac{2f'}{r}
	-\frac{1}{r^{2}}D^{2}
	\right)h_{rr}
	\nonumber\\
	&+e^{-2\psi}\left(
	\partial_{r}^{2}
	+\left(\frac{f'}{2f}-2\psi'\right)\partial_{r}
	-2\psi''
	-\frac{f'}{f}\psi'
	\right)h_{tt}
	+\frac{1}{r^{2}}\left(
	2\partial_{\theta}\partial_{r}
	+\frac{f'}{f}\partial_{\theta}
	+2\cot\theta\,\partial_{r}
	+\cot\theta\,\frac{f'}{f}
	\right)h_{r\theta}
	\nonumber\\
	&+\frac{2}{r^{2}\sin^{2}\theta}\left(
	\partial_{\phi}\partial_{r}
	+\frac{f'}{2f}\partial_{\phi}
	\right)h_{r\phi} +\frac{1}{r^{2}}\left(
	-\partial_{r}^{2}
	+\left(\frac{2}{r}-\frac{f'}{2f}\right)\partial_{r}
	-\frac{2}{r^{2}}
	+\frac{f'}{rf}
	\right)h_{\theta\theta}
	\nonumber\\
	&+\frac{1}{r^{2}\sin^{2}\theta}\left(
	-\partial_{r}^{2}
	+\left(\frac{2}{r}-\frac{f'}{2f}\right)\partial_{r}
	-\frac{2}{r^{2}}
	+\frac{f'}{rf}
	\right)h_{\phi\phi}.
	\label{Rrr_explicit}
\end{align}

\subsubsection{$\theta\theta$ component}

\begin{align}
	2\delta R_{\theta\theta} = {}&
	\left(
	\left(2f\psi'+f'+\frac{2f}{r}\right)\partial_{\theta}
	+2f\partial_{r}\partial_{\theta}
	+\frac{2f}{r}\cot\theta
	\right)h_{r\theta} +\frac{2f}{r\sin^{2}\theta}\partial_{\phi}h_{r\phi}
	+\frac{2}{r^{2}\sin^{2}\theta}\partial_{\phi}\partial_{\theta}h_{\theta\phi}
	\nonumber \\
	&+e^{-2\psi}\left(
	\partial_{\theta}^{2}
	+rf\partial_{r}
	-2rf\psi'
	\right)h_{tt}
+\Biggl (\left(\frac{f}{r}-\frac{f'}{2}-f\psi'\right)\partial_{r}
	-f\partial_{r}^{2}
	-\frac{2f}{r^{2}}
	-\frac{1}{r^{2}\sin^{2}\theta}\partial_{\phi}^{2}
	+\frac{\cot\theta}{r^{2}}\partial_{\theta}
	\Biggl) h_{\theta\theta}
	\nonumber \\
	&+\frac{1}{r^{2}\sin^{2}\theta}\left(
	-\partial_{\theta}^{2}
	+2\cot\theta\,\partial_{\theta}
	-rf\partial_{r}
	-2-2\cot^{2}\theta+2f
	\right)h_{\phi\phi}
	\nonumber \\
	&+\left(
	2rf^{2}\psi'
	-f\partial_{\theta}^{2}
	+rf^{2}\partial_{r}
	+2f^{2}
	+2rff'
	\right)h_{rr}.
    \label{Rthetatheta_explicit}
\end{align}

\subsubsection{$\phi\phi$ component}

\begin{align}
2\delta R_{\phi\phi} ={}&
\Biggl(2f\,\partial_r\partial_\phi
+\left(\frac{2f}{r} +f' +2f\psi'
\right)\partial_\phi
\Biggr)h_{r\phi}
+\frac{2}{r^2}\,\partial_\theta\partial_\phi h_{\theta\phi}
\nonumber\\
&\quad
+\sin\theta\cos\theta
\Biggl(2f\psi' +2f\partial_r +f' +\frac{2f}{r}
\Biggr)h_{r\theta}
+\frac{2f\sin^2\theta}{r}\,\partial_\theta h_{r\theta}
\nonumber\\
&\quad
+e^{-2\psi}\sin^2\theta
\Biggl(
\frac{1}{\sin^2\theta}\,\partial_\phi^2
+rf\,\partial_r
+\cot\theta\,\partial_\theta -2rf\psi'
\Biggr)h_{tt}
\nonumber\\
&\quad
+\sin^2\theta
\Biggl(2rf^2\psi' +2f^2
-\frac{f}{\sin^2\theta}\,\partial_\phi^2
+rf^2\partial_r
-f\cot\theta\,\partial_\theta
+2rff' \Biggr)h_{rr}
\nonumber\\
&\quad
+\Biggl(-f\,\partial_r^2
+\left(\frac{f}{r}
-\frac{f'}{2}
-f\psi'
\right)\partial_r
-\frac{2f}{r^2}
-\frac{1}{r^2}\,\partial_\theta^2
+\frac{2}{r^2}\cot\theta\,\partial_\theta
-\frac{2}{r^2}\cot^2\theta
\Biggr)h_{\phi\phi}
\nonumber\\
&\quad
+\frac{\sin^2\theta}{r^2}
\Biggl(-\frac{1}{\sin^2\theta}\,\partial_\phi^2
-rf\,\partial_r
+\cot\theta\,\partial_\theta
-2+2f \Biggr)h_{\theta\theta}.
\label{Rphiphi_explicit}
\end{align}

\subsection{Off-diagonal components}
\label{app:linricci_offdiagonal}

\subsubsection{$t\theta$ component}

\begin{align}
	2\delta R_{t\theta} = {}&
	\left(
	f\partial_{r}\partial_{\theta}
	+\left(f\psi'+\frac{f'}{2}\right)\partial_{\theta}
	\right)h_{tr}
	+\frac{1}{r^{2}\sin^{2}\theta}\partial_{\phi}\partial_{\theta}h_{t\phi}
	\nonumber\\
	&+\left(
	-f\partial_{r}^{2}
	+\left(f\psi'-\frac{f'}{2}\right)\partial_{r}
	-\frac{4 f}{r}\psi'
	-\frac{1}{r^{2}\sin^{2}\theta}\partial_{\phi}^{2}
	\right)h_{t\theta}.
	\label{Rthetat_explicit}
\end{align}

\subsubsection{$t\phi$ component}

\begin{align}
	2\delta R_{t\phi} = {}&
	\left(
	f\partial_{r}\partial_{\phi}
	+\left(f\psi'+\frac{f'}{2}\right)\partial_{\phi}
	\right)h_{tr}
	+\left(
	\frac{1}{r^{2}}\partial_{\theta}\partial_{\phi}
	-\frac{1}{r^{2}}\cot\theta\,\partial_{\phi}
	\right)h_{t\theta}
	\nonumber\\
	&+\left(
	-f\partial_{r}^{2}
	+\left(f\psi'-\frac{f'}{2}\right)\partial_{r}
	-\frac{4f}{r}\psi'
	-\frac{1}{r^{2}}\partial_{\theta}^{2}
	+\frac{1}{r^{2}}\cot\theta\,\partial_{\theta}
	\right)h_{t\phi}.
	\label{Rphit_explicit}
\end{align}

\subsubsection{$r\theta$ component}

\begin{align}
	2\delta R_{r\theta} = {}&
	\left(
	-\frac{2f}{r}\psi'
	-\frac{f'}{r}
	-\frac{2f}{r^{2}}
	-\frac{1}{r^{2}\sin^{2}\theta}\partial_{\phi}^{2}
	\right)h_{r\theta}
	+\frac{1}{r^{2}\sin^{2}\theta}\partial_{\phi}\partial_{\theta}h_{r\phi}
	\nonumber\\
	&+\frac{1}{r^{2}\sin^{2}\theta}\left(
	\partial_{\phi}\partial_{r}-\frac{2}{r}\partial_{\phi}
	\right)h_{\theta\phi}
	+\left(
	f\psi'\partial_{\theta}
	+\frac{f}{r}\partial_{\theta}
	\right)h_{rr} +e^{-2\psi}\left(
	\partial_{r}\partial_{\theta}
	-\left(\frac{1}{r}+\psi'\right)\partial_{\theta}
	\right)h_{tt}
	\nonumber\\
	& +\frac{1}{r^{2}}\cot\theta\left(\partial_{r}-\frac{2}{r}\right)h_{\theta\theta}
	+\frac{1}{r^{2}\sin^{2}\theta}\left(
	-\partial_{r}\partial_{\theta}
	+\cot\theta\,\partial_{r}
	+\frac{2}{r}\partial_{\theta}
	-\frac{2}{r}\cot\theta
	\right)h_{\phi\phi}.
	\label{Rthetar_explicit}
\end{align}

\subsubsection{$r\phi$ component}

\begin{align}
	2\delta R_{r\phi} = {}&
	\frac{1}{r^{3}}\left(
	r\partial_{\theta}\partial_{r}
	-2\partial_{\theta}
	+r\cot\theta\,\partial_{r}
	-2\cot\theta
	\right)h_{\theta\phi}
	\nonumber\\
	&+\frac{1}{r^{2}}\left(
	-2fr\psi'
	-\partial_{\theta}^{2}
	-2f
	+\cot\theta\,\partial_{\theta}
	-rf'
	\right)h_{r\phi}
	+\frac{1}{r^{2}}\left(
	\partial_{\theta}\partial_{\phi}
	-\cot\theta\,\partial_{\phi}
	\right)h_{r\theta}
	\nonumber\\
	&+\frac{f}{r}\left(
	r\psi'\,\partial_{\phi}+\partial_{\phi}
	\right)h_{rr}
	+\frac{1}{r}e^{-2\psi}\left(
	r\partial_{r}\partial_{\phi}-\left(r\psi'+1\right)\partial_{\phi}
	\right)h_{tt}
	+\frac{1}{r^{3}}\left(
	-r\partial_{r}\partial_{\phi}+2\partial_{\phi}
	\right)h_{\theta\theta}.
	\label{Rphir_explicit}
\end{align}

\subsubsection{$tr$ component}

\begin{align}
	2\delta R_{tr} = {}&
	\left(
	-\frac{1}{r^{2}}D^{2}
	-\frac{4f}{r}\psi'
	-2f\psi''
	-\psi' f'
	-2f(\psi')^{2}
	\right)h_{tr}
	\nonumber \\
	&+\frac{1}{r^{2}}\left(
	\partial_{\theta}\partial_{r}
	-2\psi'\partial_{\theta}
	+\cot\theta\,\partial_{r}
	-2\cot\theta\,\psi'
	\right)h_{t\theta}
	+\frac{1}{r^{2}\sin^{2}\theta}\left(
	\partial_{\phi}\partial_{r}
	-2\psi'\partial_{\phi}
	\right)h_{t\phi}.
    \label{Rtr_explicit}
\end{align}

\subsubsection{$\theta\phi$ component}

\begin{align}
	2\delta R_{\theta\phi} &=
	\left(-f\partial_{r}^{2}
	+\left( \frac{2f}{r}
	-\frac{f'}{2}
	-f\psi' \right)\partial_{r}
	-\frac{4f}{r^{2}}
	+\frac{2}{r^{2}}
	\right)h_{\theta\phi} \nonumber\\
	&\quad
	+e^{-2\psi}
	\left(\partial_{\theta}\partial_{\phi}
	-\cot\theta\,\partial_{\phi}
	\right)h_{tt} +f
	\left(-\partial_{\theta}\partial_{\phi}
	+\cot\theta\,\partial_{\phi}
	\right)h_{rr} \nonumber\\
	&\quad
	+f \left( \partial_{r}\partial_{\theta}
	+\left( \psi'
	+\frac{f'}{2f} \right)\partial_{\theta}
	-2\cot\theta\,\partial_{r}
	+\cot\theta \left( -2\psi' -\frac{f'}{f} \right) 
	\right)h_{r\phi} 
	+f \left( \partial_{r}\partial_{\phi}
	+\left( \psi' +\frac{f'}{2f}
	\right)\partial_{\phi}
	\right)h_{r\theta}.
\label{Rthetaphi_explicit}
\end{align}

\section{Angular reduction of the odd-parity field equations}
\label{app:odd_parity_angular_reduction}

This appendix records the intermediate angular manipulations used to reduce the odd-parity field equations to the radial equation \eqref{odd_parity_extr}.

\subsection{$t\theta$ equation}

Starting from \eqref{odd_Rttheta_component_simplified} and substituting
\eqref{odd_htheta_expansion}--\eqref{odd_htphi_expansion}, with $h_{t}^{lm}=h_{t}^{lm}(r)$, one obtains
\begin{align}
	2\delta R_{t\theta} &=
	\sum_{lm}\frac{1}{r^{2}\sin^{2}\theta}h_{t}^{lm}\,
	\partial_{\phi}\partial_{\theta}\Bigl(\sin\theta\,\partial_{\theta}Y_{lm}\Bigr)
	+\sum_{lm}\left(f\partial_{r}^{2}h_{t}^{lm}+\frac{4M}{r^{3}}h_{t}^{lm}\right)\frac{1}{\sin\theta}\partial_{\phi}Y_{lm} \nonumber\\
	&\quad
	+\sum_{lm}\frac{1}{r^{2}\sin^{2}\theta}h_{t}^{lm}\,\partial_{\phi}^{2}
	\left(\frac{1}{\sin\theta}\partial_{\phi}Y_{lm}\right).
\label{app_odd_Rttheta_step1}
\end{align}
The angular terms can be combined into the unit-sphere Laplacian $D^{2}$ defined in \eqref{D2_def}:
\begin{align}
	\frac{1}{\sin^{2}\theta}\partial_{\phi}\partial_{\theta}\Bigl(\sin\theta\,\partial_{\theta}Y_{lm}\Bigr)
	+\frac{1}{\sin^{2}\theta}\partial_{\phi}^{2}
	\left(\frac{1}{\sin\theta}\partial_{\phi}Y_{lm}\right)
	&=
	\frac{1}{\sin\theta}\partial_{\phi}\Bigl(D^{2}Y_{lm}\Bigr).
	\label{app_odd_Rttheta_identity}
\end{align}
Employing Eq.~\eqref{scalar_sph_eigen} then gives
\begin{align}
	\frac{1}{\sin\theta}\partial_{\phi}\Bigl(D^{2}Y_{lm}\Bigr)
	&=
	-\frac{l(l+1)}{\sin\theta}\partial_{\phi}Y_{lm}.
\end{align}
Substituting this back into Eq.~\eqref{app_odd_Rttheta_step1} yields Eq.~\eqref{odd_Rttheta_harmonic_reduction} and hence the radial equation \eqref{odd_parity_extr}.

\subsection{$t\phi$ equation}
With $h_{tr}=0$, the $t\phi$ component of $\delta R_{\mu\nu}=0$ can be written as
\begin{align}
	2\delta R_{t\phi}
	&=
	\left(\frac{1}{r^{2}}\partial_{\theta}\partial_{\phi}-\frac{\cot\theta}{r^{2}}\partial_{\phi}\right)h_{t\theta}
	+\left(-f\partial_{r}^{2}-\frac{4M}{r^{3}}-\frac{1}{r^{2}}\partial_{\theta}^{2}+\frac{\cot\theta}{r^{2}}\partial_{\theta}\right)h_{t\phi}.
	\label{app_odd_Rtphi_component}
\end{align}
Inserting \eqref{odd_htheta_expansion}-\eqref{odd_htphi_expansion} gives
\begin{align}
	2\delta R_{t\phi}
	&=
	\sum_{lm}\left(-f\partial_{r}^{2}h_{t}^{lm}-\frac{4M}{r^{3}}h_{t}^{lm}\right)\sin\theta\,\partial_{\theta}Y_{lm}
	\nonumber\\
	&\quad
	+\sum_{lm}\frac{h_{t}^{lm}}{r^{2}}
	\left(\cot\theta-\partial_{\theta}\right)
	\left(
	\partial_{\phi}\left(\frac{1}{\sin\theta}\partial_{\phi}Y_{lm}\right)
	+\partial_{\theta}\left(\sin\theta\,\partial_{\theta}Y_{lm}\right)
	\right).
	\label{app_odd_Rtphi_step1}
\end{align}
The bracket expression is related to $D^{2}$ by
\begin{align}
	\partial_{\phi}\left(\frac{1}{\sin\theta}\partial_{\phi}Y_{lm}\right)
	+\partial_{\theta}\left(\sin\theta\,\partial_{\theta}Y_{lm}\right)
	&=
	\sin\theta\,D^{2}Y_{lm}
	=
	-l(l+1)\sin\theta\,Y_{lm}.
	\label{app_odd_Rtphi_identity1}
\end{align}
Acting with $\left(\cot\theta-\partial_{\theta}\right)$ then yields
\begin{align}
	\left(\cot\theta-\partial_{\theta}\right)\left[-l(l+1)\sin\theta
	\,Y_{lm}\right]
	&=
	l(l+1)\sin\theta\,\partial_{\theta}Y_{lm}.
	\label{app_odd_Rtphi_identity2}
\end{align}
Substituting \eqref{app_odd_Rtphi_identity2} into \eqref{app_odd_Rtphi_step1} gives
\begin{align}
	2\delta R_{t\phi}
	=
	\sum_{lm}
	\left(
	-f\partial_{r}^{2}h_{t}^{lm}
	-\frac{4M}{r^{3}}h_{t}^{lm}
	+\frac{l(l+1)}{r^{2}}h_{t}^{lm}
	\right)\sin\theta\,\partial_{\theta}Y_{lm},
\end{align}
which is equivalent to \eqref{odd_parity_extr}.

\section{Identity $\delta R_{tr}=0$ in the odd-parity sector}
\label{app:odd_parity_tr_identity}

This appendix shows explicitly that the $tr$ equation is satisfied identically in the odd-parity sector under the assumptions of Sec.~\ref{subsec:exterior_perturbed}. Starting from \eqref{odd_Rtr_component} and substituting
\eqref{odd_htheta_expansion}-\eqref{odd_htphi_expansion}, with $h_{t}^{lm}=h_{t}^{lm}(r)$, one finds
\begin{align}
	2\delta R_{tr} &= \sum_{lm}\frac{1}{r^{2}}
	\left(\partial_{r}h_{t}^{lm}-2\psi' h_{t}^{lm}\right)
	\biggl(
	-\partial_{\theta}\left(\frac{1}{\sin\theta}\partial_{\phi}Y_{lm}\right)
	-\cot\theta\left(\frac{1}{\sin\theta}\partial_{\phi}Y_{lm}\right)
	+\frac{1}{\sin\theta}\partial_{\phi}\partial_{\theta}Y_{lm}
	\biggr),
	\label{app_odd_Rtr_step1}
\end{align}
where $\psi'=\partial_{r}\psi$ depends only on $r$.
The bracket vanishes identically because
\begin{align}
	\partial_{\theta}\left(\frac{1}{\sin\theta}\partial_{\phi}Y_{lm}\right)
	&= \left(\partial_{\theta}\frac{1}{\sin\theta}\right)\partial_{\phi}
	Y_{lm}
	+\frac{1}{\sin\theta}\partial_{\theta}\partial_{\phi}Y_{lm}
	= -\frac{\cot\theta}{\sin\theta}\partial_{\phi}Y_{lm}
	+\frac{1}{\sin\theta}\partial_{\phi}\partial_{\theta}Y_{lm}.
\end{align}
Therefore,
\begin{align}
	-\partial_{\theta}\left(\frac{1}{\sin\theta}\partial_{\phi}Y_{lm}\right)
	-\cot\theta\left(\frac{1}{\sin\theta}\partial_{\phi}Y_{lm}\right)
	+\frac{1}{\sin\theta}\partial_{\phi}\partial_{\theta}Y_{lm}
	&=0,
\end{align}
and \eqref{app_odd_Rtr_step1} implies $\delta R_{tr}=0$.

\section{Differentiation of $K^{lm}$ in the even-parity sector}
\label{app:evenparity_Kprime}

This appendix supplies the intermediate steps needed to differentiate Eq.~\eqref{even_K_relation} and obtain Eq.~\eqref{even_Kprime_relation}.

Starting from \eqref{even_K_relation}, we may write
\begin{align}
	K^{lm} =
	\left(\frac{1}{f}+\frac{4M}{d r f}\right)h_{tt}^{lm}
	+\frac{2M}{d f}\,\partial_{r}h_{tt}^{lm} \,.
\label{app_K_expanded}
\end{align}
Differentiating with respect to $r$ gives
\begin{align}
\partial_{r}K^{lm} =
	\partial_{r}\left(\frac{1}{f}+\frac{4M}{d r f}\right)h_{tt}^{lm}
	+\left(\frac{1}{f}+\frac{4M}{d r f}\right)\partial_{r}h_{tt}^{lm}
	+\partial_{r}\left(\frac{2M}{d f}\right)\partial_{r}h_{tt}^{lm}
	+\frac{2M}{d f}\,\partial_{r}^{2}h_{tt}^{lm} \,.
\label{app_Kprime_step1}
\end{align}
Using $f(r)=1-\frac{2M}{r}$ and $\partial_{r}f=\frac{2M}{r^{2}}$, one finds
\begin{align}
	\partial_{r}\left(\frac{1}{f}+\frac{4M}{d r f}\right)
	&= -\frac{\partial_{r}f}{f^{2}}
	-\frac{4M}{d r^{2}f}
	-\frac{4M}{d r}\frac{\partial_{r}f}{f^{2}}
	= -\frac{2M}{r^{2}f^{2}}
	-\frac{4M}{d r^{2}f}
	-\frac{8M^{2}}{d r^{3}f^{2}} \,, \\
	\partial_{r}\left(\frac{2M}{d f}\right)
	&= -\frac{2M}{d}\frac{\partial_{r}f}{f^{2}}
	= -\frac{4M^{2}}{d r^{2}f^{2}} \,.
\end{align}
Substituting these expressions into Eq.~\eqref{app_Kprime_step1} and regrouping terms yields Eq.~\eqref{even_Kprime_relation},
\begin{align}
	\partial_{r}K^{lm}
	&= \left(
	-\frac{2M l(l+1)}{d r^{2}f^{2}}
	+\left( \frac{1}{f}
	+\frac{4M(r-3M)}{d r^{2}f^{2}}
	\right)\partial_{r}
	+\frac{2M}{d f}\partial_{r}^{2}
	\right)h_{tt}^{lm},
\end{align}
where we used $l(l+1) =d+2$ in the coefficient multiplying $h_{tt}^{lm}$.

\section{Derivation of the even-parity master equation}
\label{app:evenparity_master_derivation}

Here, we outline the algebra leading from Eq.~\eqref{even_dRtt_mode}, together with Eq.~\eqref{even_Kprime_relation}, to the factorized form \eqref{even_dRtt_factored} and hence the master equation \eqref{even_master_equation}. Equation \eqref{even_dRtt_mode} with Eq.~\eqref{even_Kprime_relation} gives
\begin{align}
	0 &= \left( -f\partial_{r}^{2}
	-\frac{2f}{r}\partial_{r}
	+\frac{l(l+1)}{r^{2}}
	\right)h_{tt}^{lm} 
	+f\,\partial_{r}f
	\left(-\frac{2M l(l+1)}{d r^{2}f^{2}}h_{tt}^{lm}
	+\left( \frac{1}{f}
	+\frac{4M(r-3M)}{d r^{2}f^{2}}
	\right)\partial_{r}h_{tt}^{lm}
	+\frac{2M}{d f}\partial_{r}^{2}h_{tt}^{lm}
	\right).
\label{app_master_step1}
\end{align}
Using $\partial_{r}f=\frac{2M}{r^{2}}$ and simplifying term by term, \eqref{app_master_step1} becomes
\begin{align}
	0 &= \left(
	-f+\frac{4M^{2}}{d r^{2}}
	\right)\partial_{r}^{2}h_{tt}^{lm}
	+\left(-\frac{2f}{r}
	+\frac{2M}{r^{2}}
	+\frac{8M^{2}(r-3M)}{d f r^{4}}
	\right)\partial_{r}h_{tt}^{lm} +\left( \frac{l(l+1)}{r^{2}}
	-\frac{4M^{2}l(l+1)}{d f r^{4}}
	\right)h_{tt}^{lm} \,.
\label{app_master_step2}
\end{align}
Factoring out the common prefactor
$-\frac{1}{r^{2}}+\frac{4M^{2}}{d f r^{4}}$
from \eqref{app_master_step2} yields \eqref{even_dRtt_factored}, and hence the master equation \eqref{even_master_equation}.

\section{Reduction of the remaining even-parity field equations}
\label{app:evenparity_consistency}

This appendix summarizes how the remaining even-parity vacuum equations reduce to the same master operator as Eq.~\eqref{even_master_equation}. The derivations use the static component expressions for $\delta R_{\mu\nu}$ (Appendices~\ref{app:linricci_diagonal} and \ref{app:linricci_offdiagonal}), together with the exterior identities $e^{2\psi}=f$ and $\psi'=\frac{f'}{2f}$, the constraints \eqref{h_vac_constr}, and the harmonic identity \eqref{even_hphph_identity}.

\subsection{$rr$ equation}

After imposing the above conditions and expanding $h_{\theta\theta}=\sum_{lm}r^{2}K^{lm}Y_{lm}$, the $rr$ component reduces to
\begin{align}
	2\delta R_{rr} &=
	\sum_{lm}\Biggl(
	\left(\frac{1}{f}\partial_{r}^{2}
	+\frac{2}{f r}\partial_{r}
	+\frac{l(l+1)}{r^{2}f^{2}}
	\right)h_{tt}^{lm}
	-2\partial_{r}^{2}K^{lm}
	-\frac{2(2r-3M)}{f r^{2}}\partial_{r}K^{lm}
	\Biggr)Y_{lm} \,.
\label{app_dRrr_step1}
\end{align}
Using \eqref{even_Kprime_simple} and differentiating once gives
\begin{align}
	\partial_{r}^{2}K^{lm} &=
	\frac{1}{f}\partial_{r}^{2}h_{tt}^{lm}
	-\frac{1}{f^{2}}\partial_{r}f\,\partial_{r}h_{tt}^{lm}
	= \frac{1}{f}\partial_{r}^{2}h_{tt}^{lm}
	-\frac{2M}{f^{2}r^{2}}\partial_{r}h_{tt}^{lm} \,.
\label{app_Kpp}
\end{align}
Substituting \eqref{even_Kprime_simple} and \eqref{app_Kpp} into \eqref{app_dRrr_step1} and reorganizing produces \eqref{even_dRrr_master}.

\subsection{$\theta\theta$ and $\phi\phi$ equations}

Under the same assumptions, the angular components reduce to
\begin{align}
	2\delta R_{\theta\theta} &=
	2\left(r\partial_{r}+1\right)h_{tt}
	+\left(-f\partial_{r}^{2} -\partial_{r}f\,\partial_{r}
	-\frac{1}{r^{2}}D^{2} \right)h_{\theta\theta} \,, \\
	2\delta R_{\phi\phi} &=
	2\sin^{2}\theta\left(r\partial_{r}+1\right)h_{tt}
	+\sin^{2}\theta\left(-f\partial_{r}^{2}
	-\partial_{r}f\,\partial_{r}
	-\frac{1}{r^{2}}D^{2}\right)h_{\theta\theta} \,.
\end{align}
Substituting \eqref{even_harm_htt}-\eqref{even_harm_hthth} and using \eqref{even_Kprime_simple} reduces $2\delta R_{\theta\theta}$ to \eqref{even_dRthth_master}, and yields $2\delta R_{\phi\phi}=\sin^{2}\theta\,2\delta R_{\theta\theta}$.

\subsection{$r\theta$, $r\phi$, and $\theta\phi$ equations}
Imposing the constraints \eqref{h_vac_constr} and \eqref{even_hphph_identity} reduces the $r\theta$ and $r\phi$ components to
\begin{align}
	2\delta R_{r\theta}
	&= \frac{1}{f}\partial_{r}\partial_{\theta}h_{tt}
	+\left(\frac{2}{r^{3}}\partial_{\theta}-\frac{1}{r^{2}}\partial_{r}\partial_{\theta}\right)h_{\theta\theta} \,, \\
    2\delta R_{r\phi}
	&= \frac{1}{f}\partial_{r}\partial_{\phi}h_{tt}
	+\left(\frac{2}{r^{3}}\partial_{\phi}-\frac{1}{r^{2}}\partial_{r}\partial_{\phi}\right)h_{\theta\theta} \,.
\end{align}
After harmonic decomposition, both components are proportional to $\frac{1}{f}\partial_{r}h_{tt}^{lm}-\partial_{r}K^{lm}$ and vanish identically once \eqref{even_Kprime_simple} holds. Finally, the $\theta\phi$ component becomes \eqref{even_dRthphi_identity} and vanishes identically under $h_{rr}=f^{-2}h_{tt}$.

\section{Interior stress-energy tensor perturbation components}
\label{app:int_deltaT_components}

This appendix lists the nonvanishing components of the linearized stress-energy tensor $\delta T_{\mu\nu}$. 
\begin{align}
	\delta T_{tt} &= e^{2\psi}\delta\rho-\bar\rho\,h_{tt} \,,
	\label{app_deltaT_tt}\\
	\delta T_{rr} &= f^{-1}\delta p+\bar p\,h_{rr} \,,
	\qquad
	\delta T_{\theta\theta} = r^{2}\delta p
	+\bar p\,h_{\theta\theta}\,,
	\qquad
	\delta T_{\phi\phi} = r^{2}\sin^{2}\theta\,\delta p
	+\bar p\,h_{\phi\phi}\,,
	\label{app_deltaT_diag}\\
	\delta T_{tr} &= -\bar\rho\,h_{tr}\,, \qquad
	\delta T_{t\theta} = -\bar\rho\,h_{t\theta}\,,
	\qquad
	\delta T_{t\phi} = -\bar\rho\,h_{t\phi} \,,
	\label{app_deltaT_ti}\\
	\delta T_{r\theta} &= \bar p\,h_{r\theta} \,,
	\qquad
	\delta T_{r\phi} = \bar p\,h_{r\phi} \,,
	\qquad
	\delta T_{\theta\phi} = \bar p\,h_{\theta\phi}\,.
	\label{app_deltaT_offdiag}
\end{align}

\section{Interior component expression for $\delta R_{t\theta}$}
\label{app:int_deltaR_ttheta}

In this appendix, we record the coordinate expression for the $t\theta$ component of the linearized Ricci tensor used in the interior odd-parity analysis. Under the static constraints \eqref{int_constraints}, the $t\theta$ component of the linearized Ricci tensor can be written as
\begin{equation}
	2\delta R_{t\theta}
	= \frac{1}{r^{2}\sin^{2}\theta}\partial_{\phi}
	\partial_{\theta}h_{\phi t}+\Biggl(
	-f\partial_{r}^{2}
	+\left(f\psi'-\frac{f'}{2}\right)\partial_{r}
	-\frac{4f}{r}\psi'
	-\frac{1}{r^{2}\sin^{2}\theta}\partial_{\phi}^{2}
	\Biggr)h_{t\theta} \,.
	\label{app_int_deltaR_ttheta}
\end{equation}
Substituting the odd-parity harmonic expansions \eqref{int_odd_h_ttheta}-\eqref{int_odd_h_tphi} into Eq.~\eqref{app_int_deltaR_ttheta}, projecting onto $\partial_{\phi}Y_{lm}/\sin\theta$, and using 
Eq.~\eqref{scalar_sph_eigen} yields Eq.~\eqref{int_odd_intermediate}. The reduction of Eq.~\eqref{int_odd_intermediate} to Eq.~\eqref{int_odd_master} follows from the background identities
\begin{align}
	\psi'=\frac{m+4\pi r^{3}\bar p}{r^{2}f} \,, \qquad
	f'= -8\pi r\bar\rho+\frac{2m}{r^{2}} \,,
\end{align}
which are consequences of Eqs.~\eqref{int_background_m_f}-\eqref{int_background_psi_TOV}.

\section{Interior component expression for $\delta R_{tt}$}
\label{app:int_deltaR_tt}

This appendix details the coordinate expression for $\delta R_{tt}$ used to derive the interior even-parity equation. With $\partial_{t}h_{\mu\nu}=0$ and $h_{r\theta}=h_{r\phi}=0$, the coordinate expression used for $\delta R_{tt}$ is given by
\begin{align}
	2\delta R_{tt}
	&= f^{2}e^{2\psi}
	\Biggl( -\psi'\partial_{r} -\frac{2}{f}f'\psi'
	-2(\psi')^{2} -2\psi''
	-\frac{4}{r}\psi' \Biggr)h_{rr} \nonumber\\
	&\quad +\Biggl(
	-f\partial_{r}^{2}
	+\left(2f\psi'-\frac{f'}{2}-\frac{2f}{r}\right)\partial_{r}
	-2f(\psi')^{2} -\frac{1}{r^{2}}D^{2}
	\Biggr)h_{tt} +\frac{e^{2\psi}}{f^{2}}\psi'\left(\partial_{r}-\frac{2}{r}\right)
	\left(h_{\theta\theta}+\frac{1}{\sin^{2}\theta}h_{\phi\phi}\right).
\label{app_int_deltaR_tt_full}
\end{align}
Using $h_{\phi\phi}=\sin^{2}\theta\,h_{\theta\theta}$ and the even-parity decomposition $h_{\theta\theta}=\sum_{lm}r^{2}K^{lm}Y_{lm}$ gives
\begin{align}
	\left(\partial_{r}-\frac{2}{r}\right)h_{\theta\theta}
	= \sum_{lm}r^{2}\frac{dK^{lm}}{dr}\,Y_{lm}.
\end{align}
At this stage, we adopt the first-order relation \eqref{int_Kprime_relation},
\begin{align}
	\frac{dK^{lm}}{dr}=e^{-2\psi}\frac{dh_{tt}^{lm}}{dr},
\end{align}
as in Ref.~\cite{yang2022tidal}. This relation reduces to the standard exterior identity
$\frac{dK^{lm}}{dr}=\frac{1}{f}\frac{dh_{tt}^{lm}}{dr}$ when $e^{2\psi}=f$ holds in vacuum. Substituting the constraint $h_{rr}=f^{-1}e^{-2\psi}h_{tt}$ from Eq.~\eqref{int_constraints} into Eq.~\eqref{app_int_deltaR_tt_full}, using
Eq.~\eqref{scalar_sph_eigen},
and simplifying with the background relations \eqref{int_background_m_f}-\eqref{int_background_psi_TOV} yields Eq.~\eqref{int_deltaR_tt_simplified} quoted in the main text.

\section{Perturbed four-acceleration in the static limit}
\label{app:int_accel}

This appendix derives the expression for the perturbed four-acceleration in the static limit. The four-acceleration is defined by $a_{\mu}=u^{\sigma}\nabla_{\sigma}u_{\mu}$. In the unperturbed static configuration, $\bar a_{\mu}=\bar\Gamma^{t}{}_{t\mu}$ so that $\bar a_{\mu}=(0,\psi',0,0)$. Linearizing $a_{\mu}$ about the background four-velocity and keeping only first-order terms yields
\begin{align}
	\delta a_{\mu} =
	\delta u^{t}\bar\nabla_{t}\bar u_{\mu}
	+\bar u^{t}\bar\nabla_{t}\delta u_{\mu}
	-\frac{1}{2}\bar u^{t}\bar u^{t}\bar\nabla_{\mu}h_{tt} \,,
\end{align}
and, using Eq.~\eqref{pert_vels} together with $\partial_{t}h_{\mu\nu}=0$, this reduces to Eq.~\eqref{int_delta_a}:
\begin{align}
	\delta a_{\mu}
	= \frac{1}{2}e^{-2\psi}\left(
	0,\,
	2\psi' h_{tt}-\partial_{r}h_{tt},\,
	-\partial_{\theta}h_{tt},\,
	-\partial_{\phi}h_{tt}
	\right).
\end{align}

\section{Schwarzschild limit ($\Lambda=0$)}
\label{app:Lambda0}

In this appendix, we summarize the analytic solution of the static, even-parity sector in the asymptotically flat limit $\Lambda=0$. Setting $\Lambda=0$ in the SdS metric function \eqref{eq_f_r} gives
\begin{align}
	f(r) = 1-\frac{2M}{r}\,,
\end{align}
so the background reduces to the Schwarzschild spacetime. For each $l\ge2$, the master radial equation \eqref{master_eq} becomes
\begin{align}
	\label{Lambda0_master}	
	\left(1-\frac{2M}{r}\right) r^{2}\frac{d^{2}h}{dr^{2}}
	+2(r-3M)\frac{dh}{dr}
	-l(l+1)\,h(r)=0 \,.
\end{align}
To solve \eqref{Lambda0_master} in the exterior region $r>2M$, introduce the dimensionless coordinate
\begin{align}
	\label{Lambda0_xdef}	
	y = \frac{r}{M}-1 \,,
\end{align}
so that
\begin{align}
	\label{Lambda0_fx}
	f(r)=1-\frac{2M}{r}=\frac{y-1}{y+1}\,,
\end{align}
and derivatives transform as
\begin{align}
	\frac{d}{dr}=\frac{1}{M}\frac{d}{dy}\,, \qquad
	\frac{d^{2}}{dr^{2}}=\frac{1}{M^{2}}\frac{d^{2}}{dy^{2}}\,.
	\label{Lambda0_dtrnst}
\end{align}
Substituting \eqref{Lambda0_xdef}-\eqref{Lambda0_dtrnst} into
\eqref{Lambda0_master} yields
\begin{align}
	(y^{2}-1)\frac{d^{2}h}{dy^{2}}
	+2(y-2)\frac{dh}{dy} -l(l+1)\,h(y)=0\,.
	\label{Lambda0_hx}
\end{align}
Next, factor out a simple weight and set
\begin{align}
	h(y) = w(y)\,g(y)\,.
	\label{Lambda0_hwg}
\end{align}
Choosing $w$ so that the resulting equation for $g$ takes the standard
associated-Legendre form fixes $w$ via
\begin{align}
	\frac{w'}{w}=\frac{2}{y^{2}-1}
	\quad\Longrightarrow\quad
	w(y)=\frac{y-1}{y+1} \,.
	\label{Lambda0_w}
\end{align}
With \eqref{Lambda0_w}, Eq.~\eqref{Lambda0_hx} reduces to
\begin{align}
	(1-y^{2})\frac{d^{2}g}{dy^{2}}
	-2y\frac{dg}{dy}
	+\left(l(l+1)-\frac{4}{1-y^{2}}\right) g(y) = 0 \,,
	\label{Lambda0_assocLeg}
\end{align}
which is the associated Legendre differential equation with order $m'=2$.
Therefore,
\begin{align}
	g(y)=A_{l m}\,P_{l}^{2}(y) +B_{l m}\,Q_{l}^{2}(y)\,,
	\label{Lambda0_gsol}
\end{align}
and undoing the substitutions \eqref{Lambda0_hwg} and \eqref{Lambda0_xdef}
gives the general Schwarzschild-limit solution in the exterior region $r>2M$:
\begin{align}
	h_{tt}^{l m}(r)
	&=\left(1-\frac{2M}{r}\right)
	\Biggl(
	A_{l m}\,P_{l}^{2}\!\left(\frac{r}{M}-1\right)
	+B_{l m}\,Q_{l}^{2}\!\left(\frac{r}{M}-1\right)
	\Biggr) .
\end{align}

\end{document}